\documentclass[aps,prb,twocolumn,showpacs,preprintnumbers,amsmath,amssymb,superscriptaddress]{revtex4}

\usepackage{graphicx}
\usepackage{dcolumn}
\usepackage{color}
\usepackage{amsmath}

\makeatletter
\def\btt#1{\texttt{\@backslashchar#1}}%
\DeclareRobustCommand\bblash{\btt{\@backslashchar}}%
\makeatother

\begin{document}

\title{Quantum and thermal ionic motion, oxygen isotope effect, and 
superexchange distribution in La$_2$CuO$_4$ }

\author{P.~S.~H\"afliger}
\affiliation{Laboratory for Solid State Physics, ETH Zurich, CH-8093 Zurich, 
Switzerland}

\author{S. Gerber} 
\affiliation{Laboratory for Neutron Scattering, Paul Scherrer 
Institute, CH-5232 Villigen PSI, Switzerland}
\affiliation{Laboratory for Solid State Physics, ETH Zurich, CH-8093 Zurich, 
Switzerland}

\author{R. Pramod} 
\affiliation{Laboratory for Quantum Magnetism, \'{E}cole Polytechnique 
F\'{e}d\'{e}rale de Lausanne (EPFL), 1015 Lausanne, Switzerland}

\author{V.~I.~Schnells} 
\affiliation{Laboratory for Quantum Magnetism, \'{E}cole Polytechnique 
F\'{e}d\'{e}rale de Lausanne (EPFL), 1015 Lausanne, Switzerland}

\author{B. dalla Piazza} 
\affiliation{Laboratory for Quantum Magnetism, \'{E}cole Polytechnique 
F\'{e}d\'{e}rale de Lausanne (EPFL), 1015 Lausanne, Switzerland}

\author{R. Chati}
\affiliation{Laboratory for Solid State Physics, ETH Zurich, CH-8093 Zurich, 
Switzerland}

\author{V. Pomjakushin}
	\affiliation{Laboratory for Neutron Scattering, Paul Scherrer 
Institute, CH-5232 Villigen PSI, Switzerland}

\author{K. Conder}
\affiliation{Laboratory for Developments and Methods, Paul Scherrer Institute, 
CH-5232 Villigen PSI, Switzerland}

\author{E. Pomjakushina}
\affiliation{Laboratory for Developments and Methods, Paul Scherrer Institute, 
CH-5232 Villigen PSI, Switzerland}

\author{L.~Le~Dreau}
\affiliation{Laboratory for Developments and Methods, Paul Scherrer Institute, 
CH-5232 Villigen PSI, Switzerland}
\affiliation{Laboratory of Soft Chemistry and Reactivity, University of Rennes 
1 UMR 6226, 35042 Rennes, France}

\author{N.~B.~Christensen}
\affiliation{Department of Physics, Technical University of Denmark (DTU), 
DK-2800 Kgs. Lyngby, Denmark}
\affiliation{Laboratory for Neutron Scattering, Paul Scherrer Institute, 
CH-5232 Villigen PSI, Switzerland}

\author{O. F. Sylju{\aa}sen}
\affiliation{Department of Physics, University of Oslo, P.~O.~Box 1048 
Blindern, N-0316 Oslo, Norway}

\author{B. Normand}
\affiliation{Department of Physics, Renmin University of China, Beijing 
100872, P.~R.~China}

\author{H.~M.~R{\o}nnow}
\affiliation{Laboratory for Quantum Magnetism, \'{E}cole Polytechnique 
F\'{e}d\'{e}rale de Lausanne (EPFL), 1015 Lausanne, Switzerland}

\date{\today}

\begin{abstract}

We study the zero-point and thermal ionic motion in La$_2$CuO$_4$ by means 
of high-resolution neutron diffraction experiments. Our results demonstrate 
anisotropic motion of O and to a lesser extent of Cu ions, both consistent 
with the structure of coupled CuO$_6$ octahedra, and quantify the relative 
effects of zero-point and thermal contributions to ionic motion. By 
substitution of $^{18}$O, we find that the oxygen isotope effect on the 
lattice dimensions is small and negative ($-0.01\%$), while the isotope 
effect on the ionic displacement parameters is significant ($-6$ to 50\%). 
We use our results as input for theoretical estimates of the distribution 
of magnetic interaction parameters, $J$, in an effective one-band model 
for the cuprate plane. We find that ionic motion causes only small ($1\%$) 
effects on the average value $\langle J\rangle$, which vary with temperature 
and O isotope, but results in dramatic (10$-$20\%) fluctuations in $J$ values 
that are subject to significant (8$-$12\%) isotope effects. We demonstrate 
that this motional broadening of $J$ can have substantial effects on certain 
electronic and magnetic properties in cuprates.

\end{abstract}


\maketitle

\section{Introduction}


High-temperature superconductivity remains one of the fundamental challenges 
in condensed matter physics. More than 25 years of intensive experimental 
and theoretical studies\cite{plakida} have brought new and profound 
understanding to many branches of the physics of strongly correlated 
electrons. However, many mysteries remain concerning some basic issues 
such as the pairing mechanism, the role of the lattice, and the importance 
of structural and electronic homogeneity. Here we address perhaps the most 
basic unanswered question of all, namely ``where are the atoms ?'' It is known 
that the atoms are quantum mechanical entities, subject to a range of quantum 
fluctuations, including positional ones. Knowing the positions of the 
atoms, on time scales relevant to electronic processes, is essential to the 
understanding of any kind of model for the complex cuprate phase diagram or 
the mechanism for superconducting pairing.  


Quantum fluctuations play an essential role in the electronic and magnetic 
properties of the cuprates.\cite{kastner} In the parent antiferromagnetic 
phase, the suppressed moment\cite{huse} and the zone-boundary spin-wave 
dispersion\cite{coldea} are the fingerprints of intrinsically quantum 
mechanical effects typical of a two-dimensional (2D) $S$ = 1/2 Heisenberg 
antiferromagnet.\cite{ronnow2001} The motion of electrons in this quantum
spin medium, including their pairing tendencies, is then very strongly 
renormalized by fluctuation effects.\cite{Anderson3} The importance of 
quantum fluctuations is manifest not only in spin space but also in real 
space, in the form of zero-point motion. In C$_{60}$ materials, the effects 
of zero-point motion are found to be very significant,\cite{kohanoff} in 
the sense that they may be responsible for a substantial renormalization 
of the electron-phonon coupling and hence of superconductivity. In the 
cuprate materials, where superconducting transition temperatures ($T_c$) 
are very high, neither the effect of zero-point motion nor of thermal 
ionic motion has yet been addressed experimentally.


The isotope effect on $T_c$\cite{maxwell,reynolds} is possibly the clearest 
signature of the role of the lattice, and is one of the keys to conventional 
superconductivity. A purely phononic pairing mechanism will give a very 
characteristic dependence of superconducting properties on the mass of the 
participating ions.\cite{froehlich} In cuprates, where the contributions to 
the pairing mechanism have not been quantified, isotope effects on the 
electronic properties are well known but complex: while the pseudogap 
temperature rises\cite{petra} with $^{18}$O isotope substitution, the 
superconducting transition temperature decreases.\cite{keller} 
By contrast, only little is known regarding isotope effects on magnetic 
properties. Combined muon spin-rotation and magnetization studies\cite{rustem} 
found that the antiferromagnetic and spin-glass ordering temperatures in 
Y$_{1-x}$Pr$_x$Ba$_2$Cu$_3$O$_{7-\delta}$ exhibit a large oxygen isotope effect 
(OIE) in the regime where superconductivity and antiferromagnetic order 
coexist. A very large OIE on the spin-glass temperature has also been found 
in Mn-doped La$_{1-x}$Sr$_x$CuO$_4$ at low doping.\cite{shengelaya} These 
unusual effects could arise from the isotope-dependent mobility of the 
charge carriers.\cite{bussmann} For undoped La$_2$CuO$_4$, the N\'eel 
temperature $T_N$ was reported to decrease slightly upon oxygen 
isotope substitution,\cite{zhao} a result assumed to originate from 
structural changes.\cite{Hanzawa} To date, however, the OIE has been 
measured only for the lattice constants and the consequent orthorhombicity, 
and there is in particular no information concerning either zero-point or 
thermal ionic motion.


The electronic and magnetic properties of an interacting system depend 
fundamentally on superexchange processes between electronic orbitals, 
which mediate the electron hopping and spin-fluctuation energy scales, 
and therefore on the dimensions and geometry of the host lattice. Thus 
both zero-point (quantum) and thermal ionic motion can have significant 
consequences for the physical properties of a system, including its 
superconductivity. Quite generally, the time scale for ionic motion in 
condensed matter systems is much longer than that for electronic processes. 
However, this situation (the Born approximation) may break down for 
``low-energy'' electronic properties, especially superconductivity, in 
systems with high phonon energy scales. These observations suggest the 
importance of a careful and comparative study of ionic motion in materials 
such as cuprates. 


To investigate these open questions, we have performed a high-resolution 
neutron diffraction study of high-quality La$_2$CuO$_4$ powders. Because 
the neutron cross-section is directly proportional to the Debye-Waller 
factor, neutron diffraction is an excellent probe of zero-point and thermal 
motion. We obtain data suitable for a detailed analysis of both structural 
and thermal properties and, by using powder of high isotopic substitution,  
of the OIE on the measured quantities. Our results provide essential input 
for modelling the effects of the quantum and thermal fluctuations in ionic 
positions on the electronic and magnetic properties of the cuprate plane. 
In Sec.~II we present the details of our samples, experiments, and structural 
refinement results. Section III discusses the isotope effect and Sec.~IV 
analyzes the effective electronic models required to incorporate our measured 
motional effects into the physics of cuprates. A summary is provided in Sec.~V. 

\section{Experiment}

\subsection{Sample preparation and characterization}

Polycrystalline samples of La$_2$CuO$_4$ were prepared using conventional 
solid state synthesis. Oxygen isotope exchange was performed by annealing 
of the sample in $^{18}$O$_2$ gas (Euriso-top, 97\% isotope enrichment) at 
850 C for 30 hours.\cite{conder} The isotope content was determined by 
{\it in situ} mass spectroscopy measurements of the isotope composition 
of the O$_2$ gas in equilibrium with the sample. After the exchange process, 
the isotope enrichment was evaluated by thermal analysis, where the change 
in mass of the sample was measured during an oxygen-isotope reverse exchange 
performed in ordinary oxygen (replacing $^{18}$O by $^{16}$O). The isotope 
enrichment was found to be $78 \pm 2$\%.

The oxygen stoichiometry coefficient in both $^{16}$O and $^{18}$O samples 
was determined by thermogravimetric hydrogen reduction.\cite{conder,conder2}  
Both samples were found to have an oxygen content of $4.004 \pm 0.005$, and 
hence to be oxygen-stoichiometric within the experimental error of the 
determination procedure.

\begin{figure}[hbt]
\includegraphics[width=0.48\textwidth]{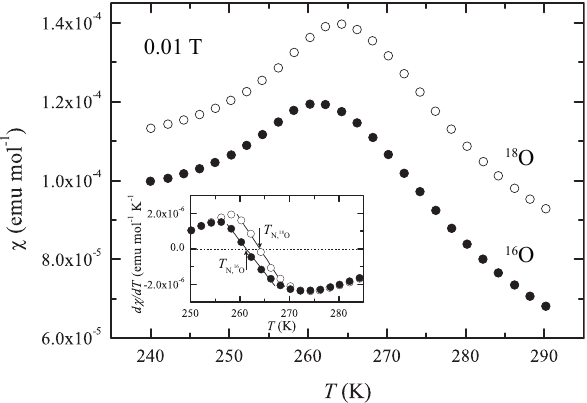}
\caption{{Susceptibility as a function of temperature for La$_2$Cu$^{16}$O$_4$ 
(solid circles) and La$_2$Cu$^{18}$O$_4$ (open circles). The inset shows the 
derivative of the susceptibility used to deduce the value of $T_N$ for both 
cases.}
\label{chi}}
\end{figure}

\subsection{Magnetization measurements}

Magnetization measurements were performed using a Quantum Design MPMS in 
fields ranging from 20~mT to 6~T at temperatures between 4 and 300~K on 
La$_2$Cu$^{16}$O$_4$ and La$_2$Cu$^{18}$O$_4$. The magnetic susceptibility, 
$\chi(T)$, obtained in an applied field of 0.1~T is shown in Fig.~\ref{chi} 
for both $^{16}$O and $^{18}$O samples. Clear peaks are observed at the onset of 
antiferromagnetic order in both cases. Several factors contribute to the 
rounding of these peaks, among which the powder nature of the sample is 
the most important. The derivatives $d\chi/dT$, shown in the inset of 
Fig.~\ref{chi}, give definitive peak values, which we take as the N\'eel 
temperatures of the two samples, $T_N = 261.4 \, \pm \, 0.1$~K for 
La$_2$Cu$^{16}$O$_4$ and $T_N = 263.8 \pm 0.1$~K for La$_2$Cu$^{18}$O$_4$. 

\begin{figure}[t]
\hspace{-10mm}\includegraphics[bb=0 0 180 220,width=0.44\textwidth]{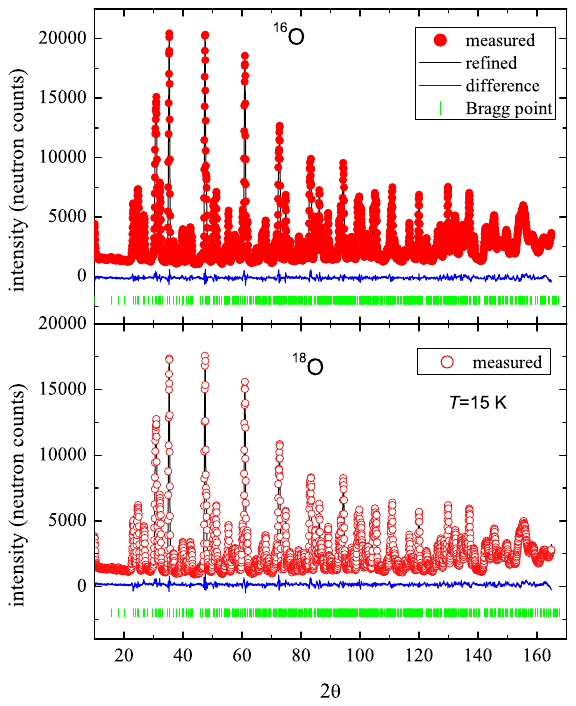}
\vspace{-10mm}
\caption{{(Color online) Neutron diffraction data at $15$ K for 
La$_2$Cu$^{16}$O$_4$ (upper panel) and La$_2$Cu$^{18}$O$_4$ (lower panel). 
The solid black line corresponds to a Rietvield refinement, and the 
difference between measured and calculated profiles is shown on the 
same scale. Tick marks below each panel represent the positions of 
allowed Bragg reflections in the $Bmba$ space group.}
\label{nd}}
\end{figure}

\begin{table*}[bt] 
\begin{tabular}{c|ccccc} 
\hline\hline
 & $a$ (\AA) & $b$ (\AA) & $c$ (\AA) & $d_{\rm Cu-O1}$ (\AA) 
& $\theta_{\rm Cu-O1-Cu}$ (deg) \\ \hline
15~K & & & & & \\
$^{16}$O & $5.33305(2)$ & $5.41783(3)$ & $13.10260(7)$ & $1.90369(5)$ &
173.434(2) \\
$^{18}$O & $5.33185(3)$ & $5.41751(3)$ & $13.09979(8)$ & $1.90343(5)$ &
173.419(2) \\ 
\hline
290~K & & & & & \\
$^{16}$O & $5.35479(3)$ & $5.40338(3)$ & $13.14810(9)$ & $1.90434(5)$ & 
174.095(2) \\
$^{18}$O & $5.35385(3)$ & $5.40283(3)$ & $13.14614(9)$ & $1.90401(5)$ & 
174.174(2) \\ 
\hline\hline
 & $y$(La) & $z$(La) & $z$(O1) & $y$(O2) &  $z$(O2) \\ \hline
15~K & & & & & \\
$^{16}$O & $-0.00839(13)$ & $0.36149(4)$ & $-0.00832(6)$ & $0.04101(12)$ 
& $0.18302(8)$ \\
$^{18}$O & $-0.00855(14)$ & $0.36158(4)$ & $-0.00834(6)$ & $0.04110(13)$ 
& $0.18293(8)$ \\ 
\hline
290~K & & & & & \\
$^{16}$O & $-0.00683(19)$ & $0.36135(4)$ & $-0.00746(8)$ & $0.03479(17)$ 
& $0.18308(9)$ \\
$^{18}$O & $-0.00655(20)$ & $0.36138(4)$ & $-0.00736(8)$ & $0.03463(17)$ 
& $0.18291(9)$ \\ 
\hline\hline
\end{tabular}
\caption{Lattice parameters, Cu--O1 bond lengths, Cu--O1--Cu bond angles, and atomic coordinates for La$_2$Cu$^{16}$O$_4$ and La$_2$Cu$^{18}$O$_4$ at temperatures of
15 and 290~K, obtained from an isotropic refinement. In the space group 
{\it Bmba} (isomorphic to {\it Cmca}, No.~64) used here, the atomic 
positions are $(0,y,z)$ for La, $(0,0,0)$ for Cu, $(1/4,1/4,z)$ for O1, 
and $(0,y,z)$ for O2.} 
\label{position}
\end{table*}

\begin{table*}[bthp]
	\centering
		\begin{tabular}{l|c|cccccc|ccccc}
& isotropic & \multicolumn{6}{c|}{half-anisotropic} 
& \multicolumn{5}{c}{fully anisotropic} \\
\hline\hline
$290~\rm K$ & $U_{\rm iso}$ & $U_{iso}$ & $U_{11}$ & $U_{22}$ & $U_{33}$ & $U_{12}$  
& $U_{23}$ &  $U_{11}$ & $U_{22}$ & $U_{33}$ & $U_{12}$  & $U_{23}$\\ \hline
$^{16}$O La & 4.6(1) & 4.8(1) & & & & & & 5.2(2) & 5.6(2) & 3.6(2) & & $-0.2(3)$ 
\\
$^{18}$O & 5.2(1) & 5.3(1) & & & & & & 5.4(2) & 6.2(2) & 4.5(2) & & $-0.4(3)$ 
\\ \hline
$^{16}$O Cu & 4.1(1) & 4.3(1) & & & & & & 1.6(3) & 3.8(3) & 7.6(3) & & 0.7(4) \\
$^{18}$O & 4.6(1) & 4.7(1) & & & & & & 2.1(3) & 4.4(3) & 8.0(4) & & 0.0(5)
\\ \hline
$^{16}$O O1 & 6.6(1) & & 4.2(2) & 4.9(3) & 11.9(4) & -1.6(2) & & 4.0(3) & 4.7(3)
 & 11.8(4) & -1.5(2) \\
$^{18}$O       & 6.0(3) & & 3.9(3) & 4.5(3) & 10.7(4) & -1.7(2) & & 3.8(3) & 
4.2(3) & 10.7(4) & -1.6(2) \\ \hline
$^{16}$O O2 & 12.3(2) & & 18.6(3) & 13.1(4) & 6.0(3) & & 0.9(4) & 18.9(4) & 
13.1(5) & 5.1(3) & & 0.4(4) \\
$^{18}$O       & 11.6(2) & & 17.7(3) & 11.7(4) & 5.9(3) & & 1.0(4) & 18.4(4) & 
11.4(5) & 4.8(3) & & 0.3(4) \\
\hline\hline
$15~\rm K$ & $U_{\rm iso}$ & $U_{iso}$ & $U_{11}$ & $U_{22}$ & $U_{33}$ & $U_{12
}$  & $U_{23}$ &  $U_{11}$ & $U_{22}$ & $U_{33}$ & $U_{12}$  & $U_{23}$\\ \hline
$^{16}$O La & 1.0(1) & 1.0(1) & & & & & & \\
$^{18}$O       & 1.5(1) & 1.5(1) & & & & & & \\ \hline
$^{16}$O Cu & 1.4(1) & 1.4(1) & & & & & & \\
$^{18}$O       & 1.8(1) & 1.9(1) & & & & & & \\ \hline
$^{16}$O O1 & 3.0(1) & & 2.0(3) & 3.6(3) & 2.9(3) & -0.4(2) & \\
$^{18}$O       & 2.3(1) & & 1.4(3) & 2.9(3) & 2.3(3) & -0.3(2) & \\ \hline
$^{16}$O O2 & 5.3(1) & & 6.2(3) & 5.4(3) & 3.9(3) & & -0.9(3) & \\
$^{18}$O       & 4.3(1) & &5.3(3) & 4.2(3) & 3.0(3) & & -0.6(3) & \\
\hline\hline
\end{tabular}
\caption{Motional parameters $U_{ij}$ for La$_2$Cu$^{16}$O$_4$ and 
La$_2$Cu$^{18}$O$_4$, expressed in units of 10$^{-3}$ \AA$^2$, as obtained 
from a fully isotropic structural refinement (left), a refinement where 
only O1 and O2 were refined anisotropically (middle), and a fully anisotropic 
refinement (right). Data for 290~K and 15~K are shown respectively in the top 
and bottom halves of the table. No convergent fit was obtained for a fully 
anisotropic fit at 15~K (see text). The shape of the symmetric $U_{ij}$ tensor 
($i,j = 1,2,3$) is given by the symmetry of the space group $Bmba$, whence 
some components are always zero; the isotropic motional parameter $U_{iso} 
= \sum_{i} U_{ii}/3$.}
\label{baniso}
\end{table*}

\subsection{Neutron diffraction}

Neutron diffraction was performed on the high-resolution powder diffractometer 
HRPT\cite{hrpt} at SINQ,\cite{sinq} located at the Paul Scherrer Institute
in Switzerland. The experiments were carried out at a wavelength $\lambda = 
1.1545$~\AA. The La$_2$Cu$^{16}$O$_4$ and La$_2$Cu$^{18}$O$_4$ samples were each 
placed in a 8~mm-diameter Vanadium container, which was mounted into a 
closed-cycle refrigerator reaching temperatures between 15 K and 290 K. 
High-statistics data were taken at 15 K and 290 K ($3.5 \times 10^7$ counts), 
whereas points at temperatures $15 < T < 290$ K were obtained with 
intermediate statistics ($2.5 \times 10^6$ counts).

The diffraction study showed that the samples of La$_2$Cu$^{16}$O$_4$ and 
La$_2$Cu$^{18}$O$_4$ both crystallized in the orthorhombic space group {\it 
Bmba} (No.~64), with atomic positions O1 $=(1/4,1/4,z)$, O2 $=(0,y,z)$, 
La $=(0,y,z)$, and Cu $=(0,0,0)$. These results are fully consistent with 
earlier reports.\cite{radaelli}

The intensity patterns obtained from neutron diffraction were refined using 
the program Fullprof,\cite{rrc} and the errors shown in Tables I-III and 
Figs.~2 and 3 are those provided by the Fullprof refinement. The diffraction 
patterns and the corresponding structural refinement of the 15~K data are 
depicted in Fig.~\ref{nd}. We emphasize that the samples were single-phased. 
The oxygen stoichiometry obtained from the structural refinement confirmed 
the results obtained by hydrogen reduction. The small excess oxygen 
concentration is consistent with the fact that rounding of the susceptibility 
peak shown in Fig.~1 is rather weak.\cite{kastner}

The ionic displacement parameters enter the refinement through the Debye-Waller 
factor, whose determination requires high data quality and good resolution 
particularly at large scattering angles. At 290~K, reliable refinements were 
achieved, which allowed us to determine a fully anisotropic set of parameters 
for the ionic motion. However, because the ionic motion is smaller at 15~K, 
its effect on the diffraction pattern was insufficiently strong for a reliable 
refinement of all 16 ionic motion parameters. Reliable refinements could be 
obtained only by constraining the motion of the Cu and La ions to be isotropic, 
while still allowing anisotropic motion of the O ions. To enable a meaningful 
comparison between the 15~K and 290~K data, we performed the same 
``half-anisotropic'' refinement at 290~K, finding the parameters extracted 
for the O ions to be very close to those given by the fully anisotropic 
refinement, and thus supporting the consistency of our results. Finally, to 
compare with previous results for the static lattice parameters, and to 
visualize the temperature dependence of the motional parameters, a set of 
refinements was also performed by restricting all of the atoms to isotropic 
displacements only. The static lattice and atomic parameters are reported in 
Table I, and were found to be completely insensitive to the choice of ionic 
motion refinement type. The ionic motion parameters extracted from all three 
types of refinement are summarized in table II. 

\begin{figure}[hbtp]
\includegraphics[width=0.372\textwidth]{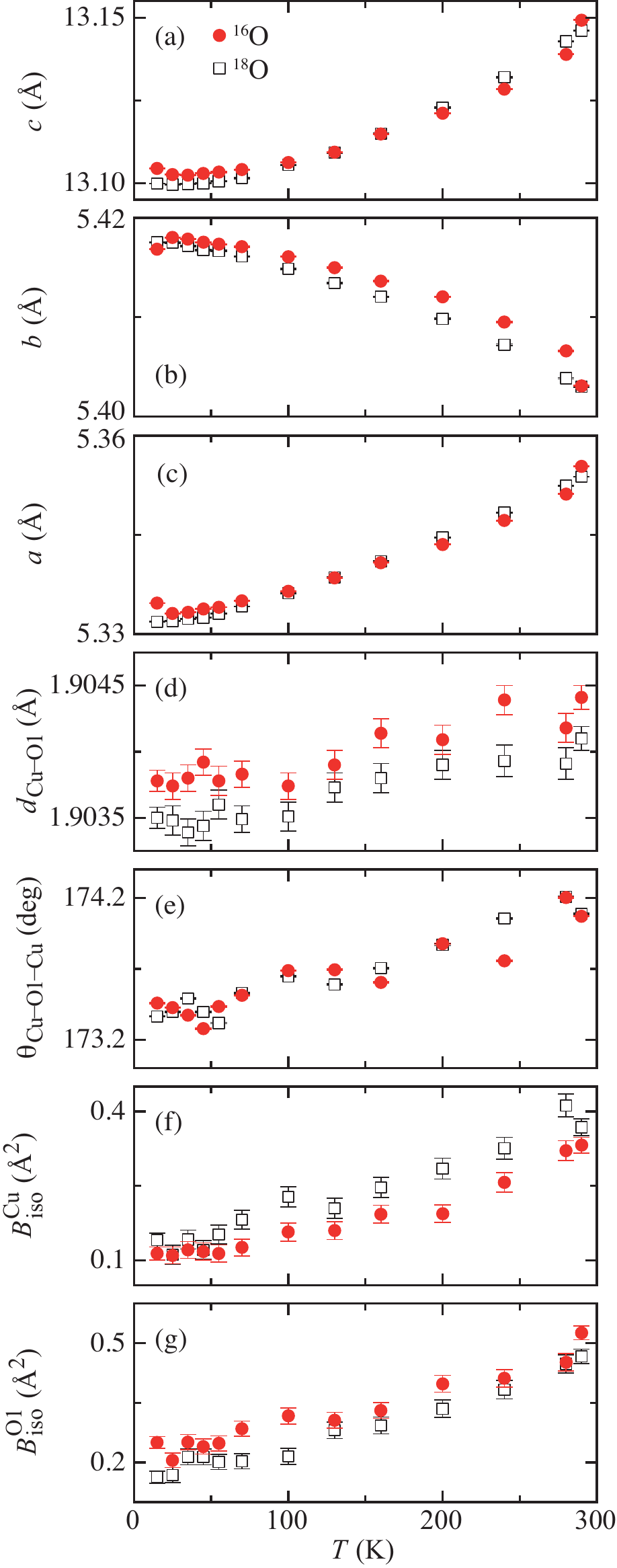}
\caption{{(Color online) Temperature dependence of structural parameters 
extracted from refinements using isotropic ionic motion parameters for 
La$_2$Cu$^{16}$O$_4$ (solid red circles) and La$_2$Cu$^{18}$O$_4$ (open black 
squares). (a-c) Lattice parameters $c$, $b$, and $a$. (d) In-plane Cu--O1 
separation, $d_{\rm Cu-O1}$. (e) Cu--O1--Cu bond angle, $\theta_{\rm Cu-O1-Cu}$. In 
addition to structural parameters, these refinements also allow the deduction 
of an isotropic motional displacement parameter, $B_{iso}$ (see text), for Cu 
ions (f) and for in-plane oxygen (O1) ions (g).}
\label{fig:Tdep}}
\end{figure}

\begin{figure*}[tbhp]
\noindent\includegraphics[width=\textwidth]{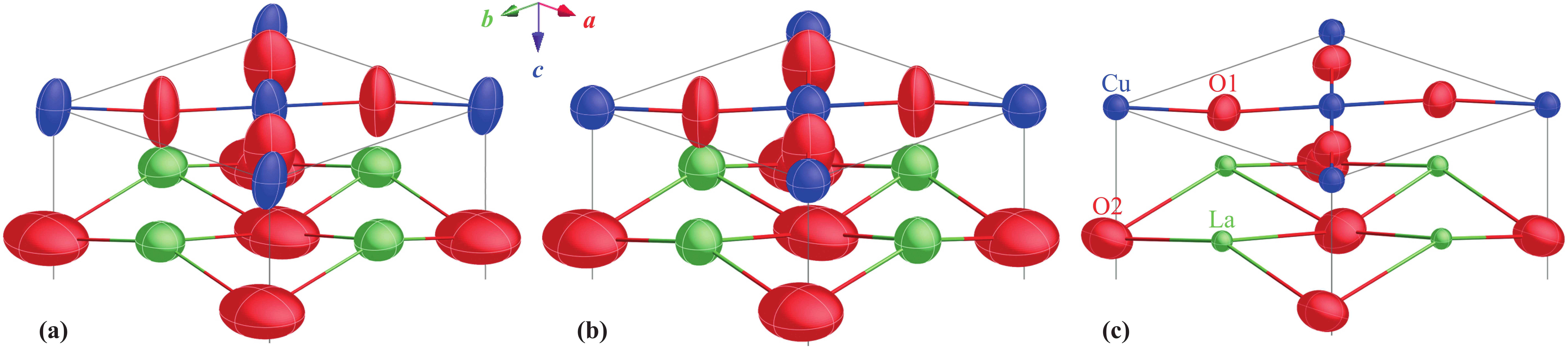}
\caption{(Color online) Crystal structure of La$_2$Cu$^{16}$O$_4$, with the 
magnitudes of the refined ionic motion parameters illustrated as spheres 
and ellipsoids. Lattice constants and motional parameters are represented on 
the same scale. La ions are shown in green, Cu in blue, and O1 and O2 in red. 
(a) Refinement of the $T = 290$~K data set with fully anisotropic thermal 
parameters. (b) Refinement of the $T = 290$~K data set performed using 
anisotropic thermal displacements for the O1 and O2 ions but isotropic 
displacements of La and Cu ions. (c) As in (b) for the $T = 15$~K data set.}  
\label{ellipsoid}
\end{figure*}


We obtain the full temperature dependence of the structural parameters, 
shown in Fig.~3, from isotropic refinements of our powder diffraction data 
at temperatures between 15 and 290 K. We remind the reader that these data 
were taken with intermediate statistics and therefore the error bars do not 
match those of Table I. Figures~3(a-c) show the lattice parameters $c$, $b$, 
and $a$, whose thermal variation is evidently rather larger than the small 
isotope effect (Sec.~III). The Cu--O1 bond lengths [Fig.~3(d)] show very 
little change, indicating that the change in lattice parameters is due 
primarily to rotation of the CuO$_6$ octahedra of which the structure is 
composed. This conclusion is reinforced by inspecting the behavior of the 
Cu--O1--Cu bond angle [Fig.~3(e)], which is the same for both in-plane bond 
directions. In Figs.~3(f) and (g) we show the isotropic motional displacement 
parameters for Cu (f) and O (g) ions, which we discuss in more detail below. 
For both ions, it can be seen that the zero-point motion and the thermal 
motion at 290~K are of similar magnitude.

\begin{figure}[h]
\centering\includegraphics[width=0.35\textwidth]{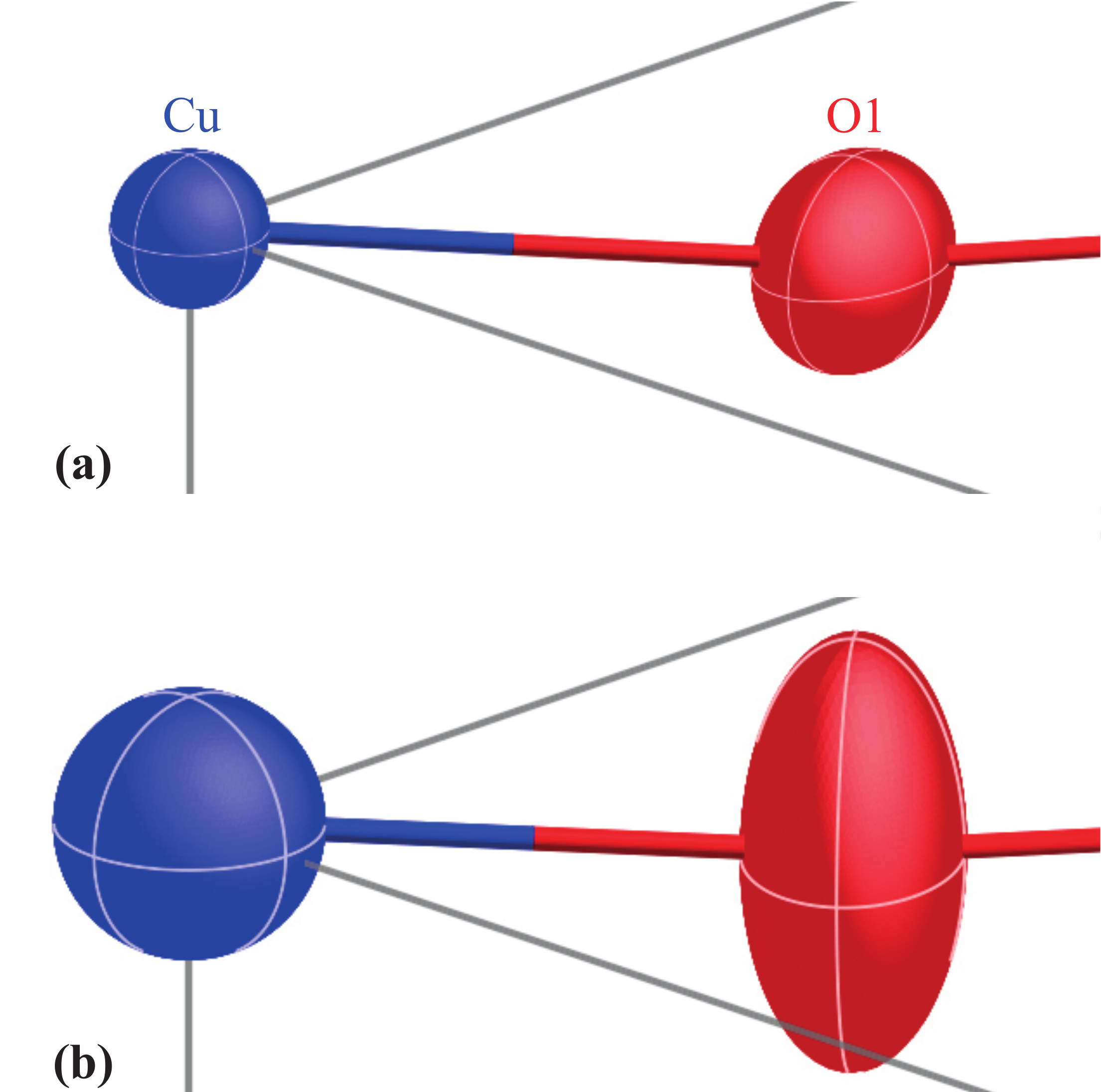}
\caption{(Color online) Zero-point and thermal contributions to ionic motion, 
compared using the parameters for Cu (blue) and O1 (red) at 15~K (a) and 290~K 
(b) given by the half-anisotropic refinement.}
\label{ellipsoid2}
\end{figure}


Turning to the details of the ionic motion contained in Table II, in 
Fig.~\ref{ellipsoid} we show the crystal structure with the ionic motion 
illustrated as spheres (isotropic) and ellipsoids (anisotropic) whose 
axes are determined by the full width at half maximum height (FWHM) of the 
ionic motion distribution, expressed by square roots of the parameters 
$U_{ij}$ in Table II. The isotropic motional parameter shown in Figs.~3(f) 
and (g) is given by $B_{iso} = 8\pi^2 U_{iso}$. The values of the anisotropic 
ionic motional parameters, represented by the ellipsoids in 
Fig.~\ref{ellipsoid}, are clearly quite different for in-plane and 
out-of-plane ions, and also within the planes. The anisotropic deformation 
of the ellipsoids of ionic motion may be understood by considering the 
nature of the various bonds in the system. In-plane oxygen atoms are 
constrained along their bond direction by the presence of Cu ions on both 
sides, causing their motion to be primarily perpendicular to the Cu--O1--Cu 
bonds. In this plane ($yz$ for an $x$-axis bond), the restoring forces are 
lowest for displacements out of the CuO$_2$ planes, and so the ellipsoids 
are most elongated in the $c$ direction. 

The same is true for the Cu ions, which are relatively tightly confined 
inside a CuO$_6$ octahedron, but with the elongation of this octahedron 
along $c$ allowing more motion in this direction. This confinement 
provides partial justification for the approximation of isotropic 
displacements necessary to refine the 15~K data, although of course this does 
not capture the effects of octahedron elongation. For the out-of-plane La and 
O2 ions, the fully anisotropic refinement at 290 K (Fig.~\ref{ellipsoid}) 
shows that their in-plane motion is stronger, and is relatively isotropic 
due to the lack of confining atoms in this structural layer. 

For all of the atoms, zero-point fluctuations account quite uniformly for 
rather more than half of the net ionic motion at 290 K (Fig.~\ref{ellipsoid}).
This is quantified for the in-plane ions in Fig.~\ref{ellipsoid2}, which 
compares the ionic motion of the Cu and O1 ions at 15~K and 290~K. At 15~K, 
the zero-point motion of O1 is almost isotropic in the $y$- and $z$-directions, 
but the motion along $z$ grows significantly with temperature, causing the 
distortion of the motional ellipsoids we observe at 290 K. As a consequence, 
the leading effect of thermal motion on the Cu--O1--Cu bonds is on the bond 
angle [Fig.~3(e)]. In fact this observation explains the asymmetric 
distribution of magnetic interaction strengths at high temperature, 
which we will discuss in Sec.~IV.

\begin{table*}[tb]
\begin{tabular}{c|ccc|cc|cccc}
&  \multicolumn{3}{c|}{Lattice parameters} & \multicolumn{2}{c|}{Bond 
parameters} & \multicolumn{4}{c}{$\Delta B_\mathrm{iso}$} \\
\hline\hline
$T$ & $\Delta a$ $[10^{-4}]$  & $\Delta b$ $[10^{-4}]$ & $\Delta c$ $[10^{-4}]$ 
& $\Delta d_{\rm Cu-O1}$ $[10^{-4}]$ & $\Delta(180 - \theta_{\mathrm{Cu-O1-Cu}})$& 
La & Cu & O1 & O2\\ \hline
15~K & $-2.25(7)$ & $-0.59(8)$ & $-2.14(8)$  & $-1.37(37)$ & 0.0023(4) & 
0.50(18) & 0.29(12) & $-$0.23(4) & $-$0.19(2) \\
290~K & $-1.76(8)$ & $-1.02(8)$& $-1.49(10)$ &  $-1.73(37)$ & $-$0.0134(5) & 
0.13(3) & 0.12(4) & $-$0.09(2) & $-$0.06(2)
\\ \hline\hline
\end{tabular}
\caption{OIE on the lattice parameters, on the Cu--O1--Cu bond distance and 
angle, and on the isotropic ionic motion parameters in La$_2$CuO$_4$.}
\label{lattice_oie}
\end{table*}

\section{Oxygen Isotope Effect}

We express the OIE on a quantity $x$, by $\Delta x = \frac{x^{18} - x^{16}}
{x^{16}}$, where $x^{16}$ and $x^{18}$ are shorthand for the quantity $x$ 
measured respectively in La$_2$Cu$^{16}$O$_4$ and La$_2$Cu$^{18}$O$_4$. We 
begin our analysis of the OIE in La$_2$CuO$_4$ by considering the 
susceptibility (Fig.~1). Oxygen isotope substitution raises the 
susceptibility peak by $2.4 \pm 0.2$~K, and hence the OIE $\Delta T_N 
= 0.92 \pm 0.07\%$ is positive. While this result contradicts that 
reported in Ref.~\onlinecite{zhao}, we emphasize that these authors 
found their values of $T_N$, and thus also of the OIE on $T_N$, to 
depend strongly on the preparation and annealing conditions of their 
samples.  

Turning to the structural analysis, the OIE on the ionic position and 
motion can be extracted from the information in Table II. The OIEs obtained 
for the lattice parameters $a$, $b$, and $c$, the Cu--O1 bond length, the 
Cu--O1--Cu bond angle, and the isotropic ionic motion parameters at 15~K 
and at 290~K are summarized in Table~\ref{lattice_oie}. The temperature 
dependence of each of these parameters is shown in Fig.~\ref{fig:Tdep}. 

The OIE on the crystal lattice corresponds to a very small overall contraction 
of order $10^{-4}$. The negative OIE on the $c$-axis lattice parameter, which 
is expected to increase the weak interactions in this direction, agrees 
qualitatively with the positive OIE we measured for the magnetic ordering 
temperature $T_N$. The OIE on the Cu--O1 bond length is also negative, which 
will have a minor effect on the in-plane interaction parameters we discuss in 
Sec.~IV. The OIE on the static bond angle is entirely negligible. 

\begin{figure}[h]
\centering\includegraphics[width=0.48\textwidth]{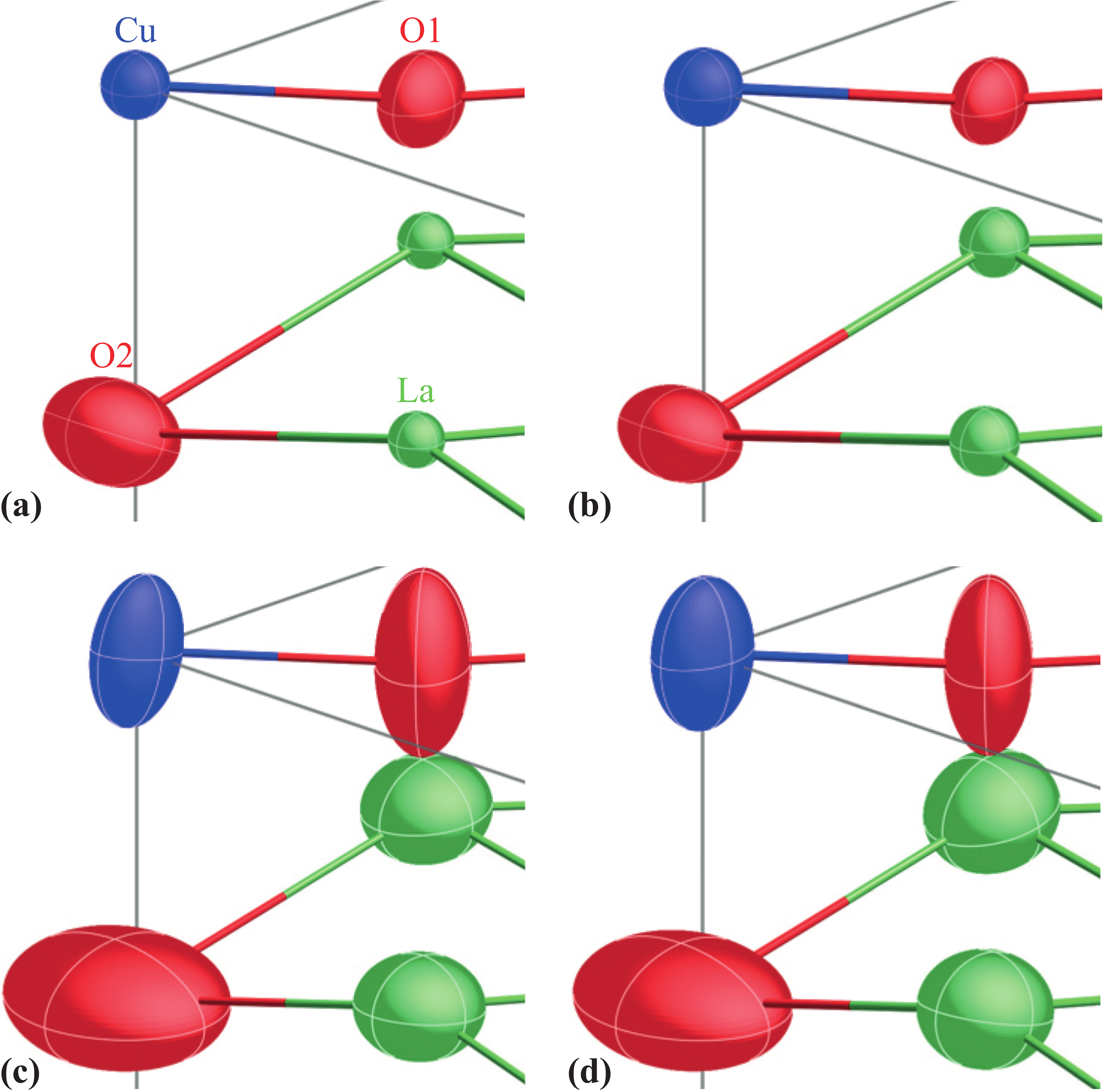}
\caption{(Color online) Oxygen isotope effect on the ionic motion parameters 
from comparison of La$_2$Cu$^{16}$O$_4$ (a,c) and La$_2$Cu$^{18}$O$_4$ (b,d). 
(a,b) Results from the half-anisotropic structural refinement at 15~K show 
a reduction in O1 and O2 motion leading to a corresponding increase in 
Cu and La motion. (c,d) Results from the fully anisotropic structural 
refinement at 290~K show that thermal fluctuations suppress the OIE on 
the motional parameters while increasing their $c$-axis anisotropy.} 
\label{ellipsoid3}
\end{figure}

However, the OIE on the ionic motion is significant, ranging from 
$-23\%$ to $+50\%$ in some parameters. The motional ellipsoids of all 
the ions for La$_2$Cu$^{16}$O$_4$ and La$_2$Cu$^{18}$O$_4$ are compared in 
Fig.~\ref{ellipsoid3}. At low temperatures, a careful inspection of the 
half-anisotropic refinements for the two samples [Figs.~\ref{ellipsoid3}(a) 
and (b)] shows a reduction, or negative OIE, of order 20\% in zero-point O1 
and O2 ionic motion from $^{16}O$ to $^{18}O$, and that this is accompanied 
by a similar positive OIE in Cu and La motion (Table III). At 290 K, the 
fully anisotropic refinements in Figs.~\ref{ellipsoid3}(c) and (d) show 
that the thermal contributions to ionic motion are similar for both $^{16}O$ 
to $^{18}O$, effectively suppressing the motional OIEs to values of order 
10\% (Table III). The increasing thermal motion is also quite anisotropic, 
displaying strong increases in the $c$-axis motion of the in-plane Cu and 
O1 ions and a corresponding flattening in the ellipsoids of the out-of-plane 
La and O2 ions. These results are qualitatively consistent with expectations 
from a simple ball-and-spring model for the structure of anisotropic CuO$_6$ 
octahedra and with the sum rule for lattice vibrations.\cite{rosenstock} 
Isotopic substitution also has a small effect on the directions of ionic 
motion, which can be seen in the orientation of the ellipsoids in 
Figs.~\ref{ellipsoid3}(c) and (d). 

\section{Motional renormalization of electronic and magnetic properties}

Changes in ionic positions have a natural effect in altering the 
electronic and magnetic coupling constants. In the effective one-band, 
strong-coupling model for the behavior of the doped cuprate plane in 
La$_2$CuO$_4$, these are denoted as $t$ for the hopping of hole-like 
quasiparticles and $J$ for the antiferromagnetic superexchange interaction 
between $S = 1/2$ spins.\cite{rzr} Here we focus primarily on $J$, and 
estimate the effects of ionic motion by following detailed theoretical 
studies of a single cuprate plane.\cite{rej} 

The energy of thermal motion in La$_2$CuO$_4$ lies largely in the range 
10$-$20 meV,\cite{boeni} while the bandwidth of antiferromagnetic exchange 
processes extends up to 300 meV,\cite{coldea} corresponding to $J \simeq 
140$ meV. Thus the ionic displacement is slow compared to the time scale 
(inverse energy scale) of the electronic parameters and the Born approximation 
may be justified. Here we comment that an alternative means of incorporating 
lattice effects on the electronic and magnetic properties would be a 
first-principles lattice dynamics calculation for the complete phonon 
spectrum.\cite{rld} From this one may determine the specific phonons most 
important for particular electronic coupling effects; these are usually 
thought to be the ``breathing modes'' of the cuprate squares, involving 
longitudinal motion of Cu and O ions along their bonding axes. However, 
phonons relevant for this type of process, which do have significant 
effects on the quasiparticle properties, usually lie in the 50$-$80 meV 
region\cite{phononenergy} and are not important at room temperature.

\subsection{Estimation of superexchange parameters}

The effects of temperature on magnetic interactions have been considered 
in a general microscopic framework,\cite{rb} where the dominant behavior 
is a sharp fall in $J(T)$ due to thermal expansion of the system. This 
effect, which in some systems is large enough to be observable in the 
susceptibility peak position, is contained in our thermal data but turns 
out to be weak. We begin a more specific microscopic analysis by considering 
the integral describing the overlap of Cu $d$- and O1 $p$-orbitals. From 
the general theory of orbital overlap,\cite{rhb} 
\begin{equation}
t_{pd} = A_0 d^{-\alpha_0} \cos^{\beta_0} \theta 
\label{etpd}
\end{equation}
where $d = |{\bf r}_{\rm Cu} - {\bf r}_{\rm O}|$ is the spatial separation
of the Cu and O ions and $\theta$ describes their angular deviation away 
from the situation of a perfect $\sigma$-orbital alignment; thus $d \equiv 
d_{\rm Cu-O1}$ and $\theta \equiv \theta_{\rm Cu-O1-Cu}$ are precisely the structural 
parameters of Sec.~II. $A_0$ is a constant of proportionality and the 
power-law dependences in $d$ and $\cos \theta$ are given respectively 
by $\alpha_0 = 3$ or $3.5$ and $\beta_0 = 1$. 

The lowest-order processes in a perturbative expansion for the quasiparticle 
hopping and superexchange interaction yield the parameter dependences $t 
\propto t_{pd}^2$ (second order) and $J \propto t_{pd}^4$ (fourth order). 
However, the cuprate plane is rather poorly described by perturbative 
approaches, and a more detailed analysis\cite{rej} reveals the importance 
of direct in-plane O--O hopping, described by the overlap integral $t_{pp}$, 
which contributes to $J$ through the many possible fifth-order processes. To 
take account of such contributions, we formulate the problem by computing 
the effective overlap integral
\begin{equation}
J = A (d_1 d_2)^{-\alpha/2} (\cos \theta_1 \cos\theta_2)^{\beta/2},
\label{ej}
\end{equation} 
for a given bond Cu(1)--O--Cu(2). Here $d_1 = |{\bf r}_{\rm Cu(1)} - 
{\bf r}_{\rm O}|$ and $d_2 = |{\bf r}_{\rm Cu(2)} - {\bf r}_{\rm O}|$ are the 
respective separations of the two Cu ions from the same O ion, while 
$\theta_1$ and $\theta_2$ are the corresponding bonding angles and $\alpha$ 
and $\beta$ denote effective power-law dependences. By considering only the 
lowest-order contribution to $J$, one would expect the very strong powers 
$\alpha \simeq 12-14$ and $\beta = 4$. However, the extensive contributions 
from higher-order terms,\cite{rej} primarily at 5th and 6th order in $t_{pd}$ 
and $t_{pp}$, lead to effective powers closer to $\alpha = 7$ and $\beta = 2$. 
The best theoretical estimate provided by Ref.~\onlinecite{rej} was $J \propto 
d^{-6.9}$, which was in good agreement with an experimental estimate $J \propto 
d^{-6.4}$ deduced\cite{raea} from a high-pressure study. 

Here we use the form of Eq.~(\ref{ej}) in combination with a statistical
distribution of Cu and O positions whose probabilities are determined from 
the experimental measurements represented in Fig.~\ref{ellipsoid}. Mindful 
of the fact that the CuO$_2$ lattice contains two O atoms for each Cu, we 
calculate superexchange parameters not for a single bond (two Cu and one 
O atom) but for a small lattice, in order to represent appropriately the 
weight of each probabilistic function. Our primary analysis of the resulting 
data is simply to take a histogram for the probability of finding a given 
bond strength, and this takes into account all of the correlations between 
the positions of two ions in one bond, including the thermal-expansion 
effect.\cite{rb} However, a secondary effect is that the motion of one 
Cu ion clearly induces correlations among all the neighboring bonds 
(specifically, $\langle i-x,i \rangle$, $\langle i,i+x \rangle$, $\langle 
i-y,i \rangle$ and $\langle i,i+y \rangle$), and we comment on this point 
below.

\begin{figure}[t]
\noindent\includegraphics[scale=0.2]{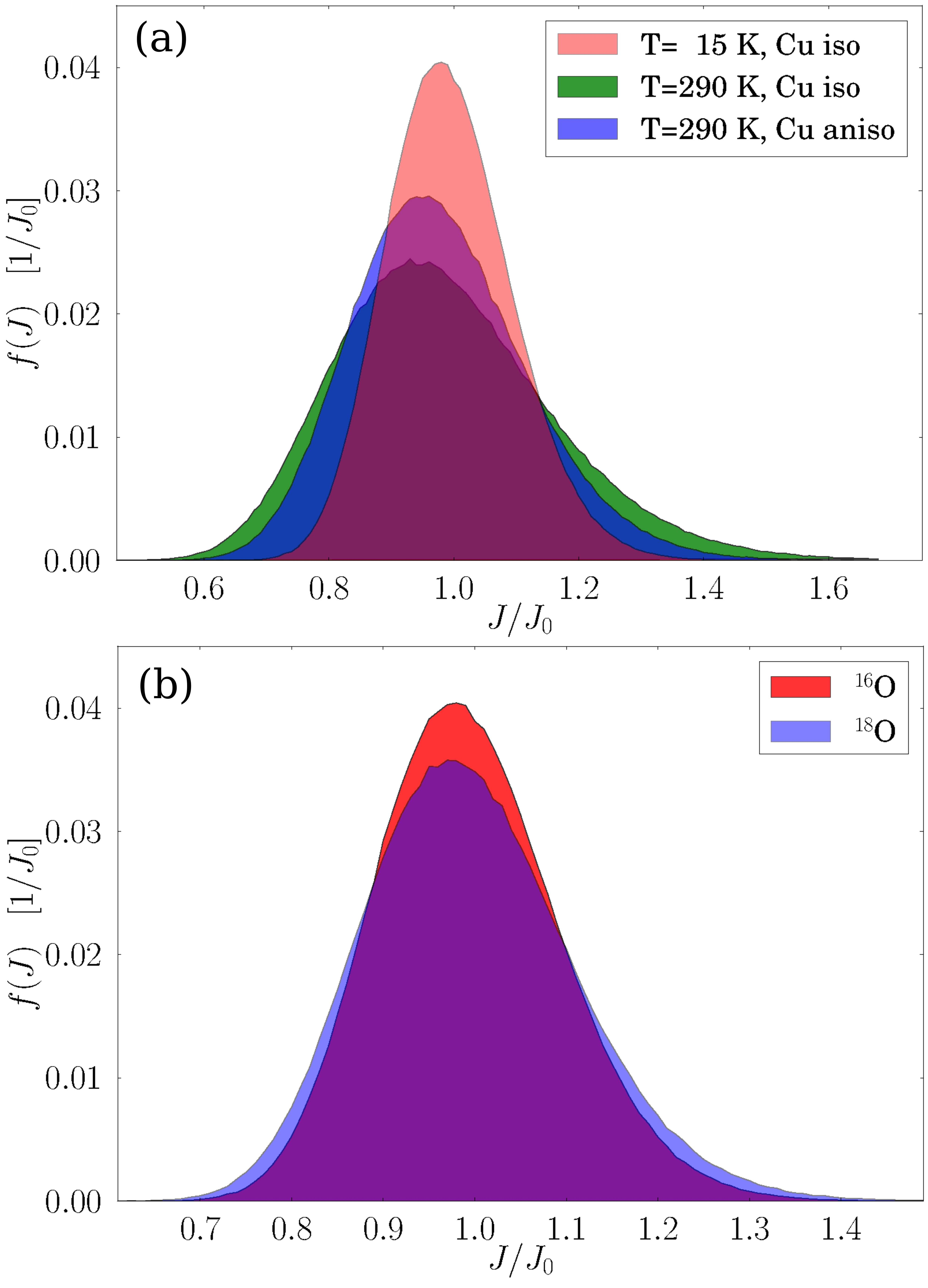}
\caption{(Color online) Distribution histogram of the superexchange parameter 
$J$, normalized to its uniform-lattice value $J_0$ (Table IV), due to 
zero-point and thermal Cu and O ionic motion. (a) Histograms for the three 
structural refinements of La$_2$Cu$^{16}$O$_4$. (b) Histograms comparing $J$ 
for La$_2$Cu$^{16}$O$_4$ and La$_2$Cu$^{18}$O$_4$ at 15~K with half-anisotropic 
motional parameters.}
\label{histograms}
\end{figure}

\subsection{Motional renormalization}

In Fig.~\ref{histograms}(a) we show the distribution functions obtained for 
the superexchange parameter $J$ in La$_2$Cu$^{16}$O$_4$ at the two experimental 
temperatures ($T = 15$ K and 290 K) and in Fig.~\ref{histograms}(b) we compare 
the distributions for La$_2$Cu$^{16}$O$_4$ and La$_2$Cu$^{18}$O$_4$ at 15~K.
Qualitatively, it is clear that there is a significant distribution of $J$ 
values even at low temperature. The increase in width of the distribution 
from 15 K to 290 K is a consequence of the increase in $U$ values shown in 
Table II and Fig.~\ref{ellipsoid}, and will be quantified below. The width 
of the distribution at 290~K differs slightly between the half-anisotropic 
and fully anisotropic refinements, but the qualitative features are the same 
[Fig.~\ref{histograms}(a)]. The distribution is slightly wider for 
La$_2$Cu$^{18}$O$_4$ than for La$_2$Cu$^{16}$O$_4$ [Fig.~\ref{histograms}(b)], 
indicating that the effects on $J$ of the reduced O ion motion 
[Fig.~\ref{ellipsoid3}] are more than compensated by the additional Cu ion 
motion this allows. It is also evident that the distributions are asymmetric, 
with the maximum value $J_\mathrm{m}$ shifting down in energy but a longer tail 
extending towards higher values of $J$. We discuss the possible consequences 
of such a distribution for the magnetic and also the electronic properties 
of the cuprates below.

\begin{table}[tbh]
\begin{tabular}{l|cccc}
& $J_0/J_0^r$ & $\langle J \rangle/J_0^r$ & $J_{\mathrm{m}}/J_0^r$ & $\Delta 
J/J_0^r$ \\ 
\hline\hline
$^{16}$O 15~K half-anis.   &    1.000  &   0.999 &  0.984 & 0.099 \\
$^{18}$O 15~K half-anis.   &    1.001  &   1.002 &  0.982 & 0.111 \\
$^{16}$O 290~K half-anis.  &    0.995  &   0.992 &  0.947 & 0.163 \\
$^{18}$O 290~K half-anis.  &    0.996  &   0.995 &  0.945 & 0.171 \\
$^{16}$O 290~K full anis.  &    0.995  &   0.983 &  0.956 & 0.133 \\
$^{18}$O 290~K full anis.  &    0.996  &   0.987 &  0.953 & 0.144 \\ \hline\hline
\end{tabular}
\caption{Analysis of superexchange parameters determined from the ionic motion 
contained in the six structural refinements. The ``static'' value $J_0$ is 
calculated for the ionic positions at the centers of the distributions, and 
is normalized to the value $J_0^r$ for La$_2$Cu$^{16}$O$_4$ at 15~K. The average 
superexchange value $\langle J \rangle$ is taken over the whole distribution.
The most likely value is denoted as $J_\textrm{m}$. The standard deviation 
of the distribution is specified by $\Delta J$, with the FWHM given by 
$2.35 \Delta J$.}
\label{tab:Js}
\end{table}

First we consider the properties of the histograms we compute. Table 
\ref{tab:Js} characterizes the histograms for the six different structural 
refinements by their mean, their peak position (corresponding to the most 
likely value of $J$), and their standard deviation. The parameter $J_0$ is 
the superexchange interaction expected if the ions are stationary and 
located in their conventional atomic positions. All of the values quoted 
for $\langle J \rangle$, $J_\textrm{m}$, and $\Delta J$ in Table \ref{tab:Js} 
are normalized to $J_0^r$, calculated for La$_2$$^{16}$CuO$_4$ at 15 K.

The variation we compute in the static value $J_0$ is as expected. It is 
marginally higher for $^{18}$O due to the slight contraction of the lattice, 
and marginally lower at 290~K due to the thermal expansion of the lattice. 
The variations are, however, tiny and the $0.1\%$ OIE is insufficient to 
account for the $0.9\%$ increase in $T_N$. Including the effects of ionic 
motion leads to a variation larger by a factor of 3$-$4 as a function of 
isotope and temperature, and the $0.4\%$ OIE on $\langle J\rangle$ is 
a much more significant factor in explaining the change in $T_N$. Here we 
remind the reader that $T_N$ is a combination of the in-plane correlation 
length, which is controlled by $J$ and also increases exponentially with 
decreasing temperature,\cite{ronnow1999} and the small interlayer coupling, 
which is more difficult to estimate, but is expected to increase as the 
$c$-axis contracts. We return to this topic below.

The most remarkable result of our study is unquestionably the 
significant standard deviation of the distribution, which we find to vary 
between 10 and 20\% for realistic parameters in Eq.~(\ref{ej}). Further, 
while the OIE on the average value $\langle J \rangle$ is only $0.4\%$, 
the widths of the $J$ distributions are 8$-$12\% larger for $^{18}$O than 
for $^{16}$O. Experimental observables sensitive to this width could therefore 
show a significant OIE. The standard deviation $\Delta J$ quantifies the 
observation made in Fig.~\ref{histograms}(a), that the width of the $J$ 
distribution increases by a factor of approximately 1.6 on increasing the 
temperature from 15 to 290 K. This indicates that thermal fluctuations 
contribute rather less to the effect of ionic motion than do the zero-point 
fluctuations even at room temperature, and thus that our considerations are 
important at all temperatures in cuprates. We comment also that the width of 
the distribution resulting from the fully anisotropic structural refinement 
is significantly smaller than for the half-anisotropic one, and we expect 
that this is a consequence of the enhanced $z$-axis Cu motion in the former 
fit (Table II) appearing as in-plane motion in the latter.

\begin{figure}[tb]
\noindent\hspace{-2mm}\includegraphics[width=0.49\textwidth]{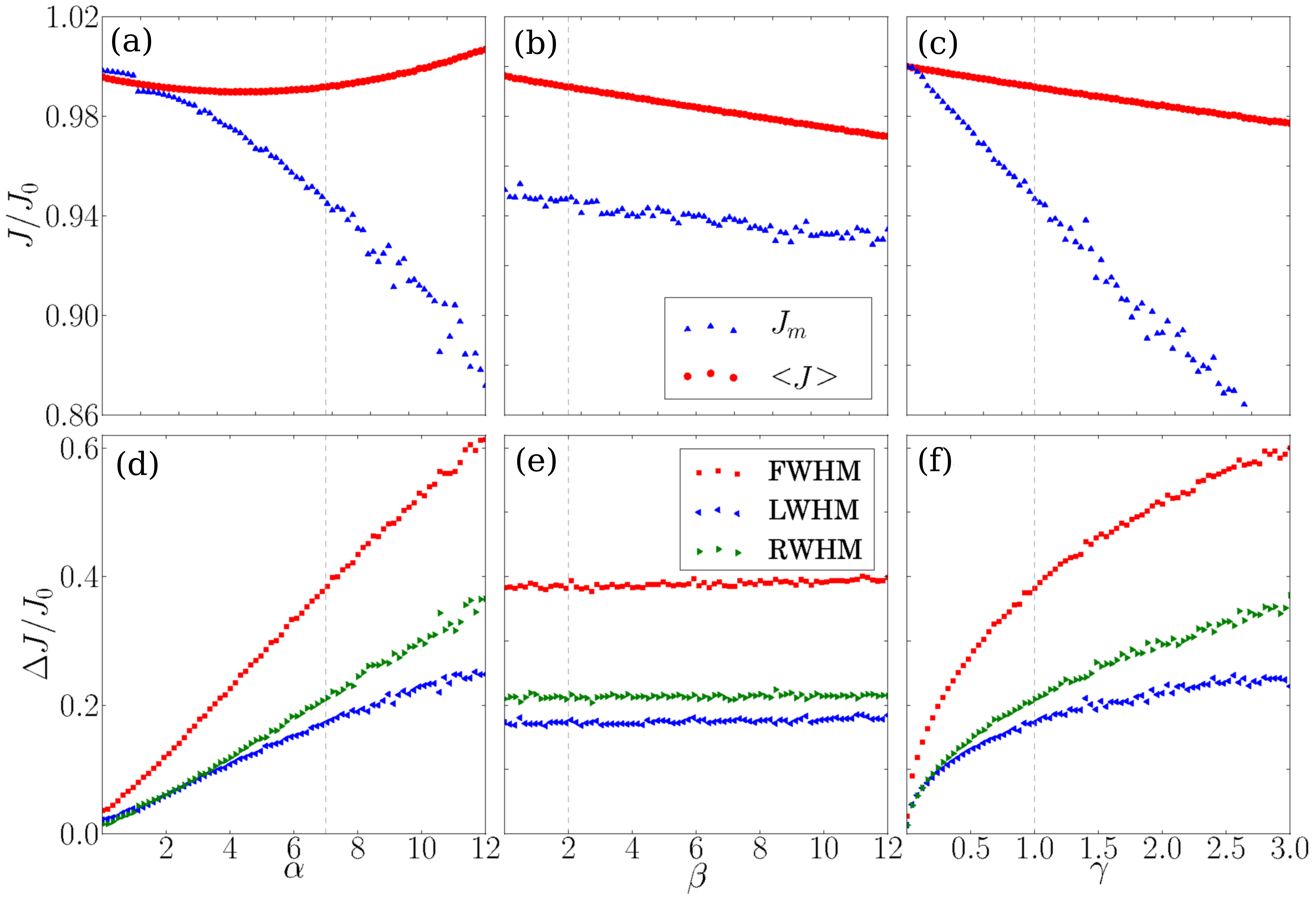}
\caption{(Color online) Variation of the superexchange parameter ($J$) 
distribution with (a,d) the exponent $\alpha$ governing the dependence on 
ionic separation $d$ [Eq.~(\ref{etpd})], (b,e) the exponent $\beta$ governing 
the dependence on relative orbital orientation $\theta$ [Eq.~(\ref{etpd})], 
and (c,f) the scale factor $\gamma$ applied to the ionic motion distribution 
(an analog of the temperature). Panels (a), (b), and (c) show the average 
$\langle J\rangle$ and peak value $J_\mathrm{m}$, panels (d), (e), and (f) 
the left-width, right-width, and full-width at half maximum $\Delta J$ 
of the distribution.}
\label{exponents}
\end{figure}

Considering briefly the functional dependence of the superexchange 
distribution, the average changes little while the peak value falls 
significantly with the power $\alpha$ governing the ionic separation 
[Fig.~8(a)], whereas both fall only weakly with the power $\beta$ 
governing the angle [Fig.~8(b)]. By contrast, the standard deviation 
increases linearly with $\alpha$ [Fig.~8(d)] but is essentially 
independent of $\beta$ [Fig.~8(e)]. These results are fully in line 
with physical expectations: a linear increase in the standard deviation 
is, for a Gaussian, the response to an increase in the power of the 
function with which the distribution falls away from its mean; because 
the Cu and O ions are moving close to an angle of 180$^0$, the lowest-order 
effect of an angular deviation is small and is not affected by a change in 
$\beta$. In evaluating the mean and standard deviation for the distributions 
in Table \ref{tab:Js}, we have taken the values $\alpha = 7$ and $\beta = 
2$.\cite{rej} While we do not have the experimental data to analyze all 
temperatures, the effect of temperature may be simulated simply by 
applying a scale factor to the ionic motion parameters. The average 
$\langle J \rangle$ and peak value $J_\mathrm{m}$ decrease linearly with the 
amplitudes of ionic motion [Fig.~8(c)], while the standard deviation $\Delta 
J$ increases approximately as a square root of this amplitude [Fig.~8(f)]. 
This last result explains why the approximate doubling of ionic motion 
amplitudes between 15 K and room temperature (Table II) leads only to a 
$60\%$ broadening in the distribution of $J$ values (Table IV).

\subsection{Consequences of motional renormalization}

We turn now to a discussion of the effects on the electronic and magnetic 
properties in the cuprate materials of the fact that $t$ and $J$ do not 
have fixed values. Instead, both parameters obey a fluctuating distribution 
of values whose mean drops slightly with increasing temperature due to the 
thermal expansion of the lattice,\cite{rb} and whose functional form is an 
approximately Gaussian distribution where the width is determined by both 
zero-point and thermal fluctuations. The width of the distribution is 
significant even at the lowest temperatures, spanning a range of order 
30 meV (Fig.~7). While the thermal expansion is, from our data, small in 
cuprates and thus has little effect on the mean value $\langle J \rangle$, 
the broadening $\Delta J$ may have important effects on thermodynamic 
measurements, such as the susceptibility, and on dynamical measurements 
such as the optical response and the magnon spectrum. These latter effects 
can be expected primarily in the widths of the excited modes, rather than 
in their positions, but may also induce interactions between excitations. 

A key question is how ionic motion may affect the N\'eel temperature. Our 
analysis gives only a partial impression of thermally induced shifts in 
$T_N$, because this quantity is critically dependent on the $z$-axis coupling 
and the superexchange for this direction cannot be estimated accurately from 
our data. For doped cuprates, one of the most important sets of static and 
dynamical effects arising from motional broadening of $t$ and $J$ will be 
on superconductivity itself. Here only our zero-point motional results are 
relevant, and their effects can include changes to $T_c$, to the gap 
$\Delta({\bf k})$, and also to the origin of an isotope effect. While $T_c$ 
is a consequence of three-dimensional coupling, and thus is subject to some 
of the same factors as $T_N$, $\Delta({\bf k})$ is a largely in-plane quantity 
thought in many theories to depend linearly on $J$, and would therefore be 
reduced and broadened in quantities such as the quasiparticle dispersion. 
Our experimental results show a very weak OIE on the average in-plane 
superexchange interaction. However, motional broadening is 8$-$12\% larger 
in the $^{18}$O system than in the $^{16}$O material, and this will affect 
the quasiparticle properties. Thus our results suggest that the origin of 
the isotope effects measured for a variety of static and dynamic quantities 
in cuprates,\cite{rie1} including the infra-red optical response\cite{rie2} 
and the photoemission spectrum,\cite{rie3} lies in the $z$-axis coupling 
(above) for static quantities and in the nature of the quasiparticles for 
dynamical ones. 

To quantify some of the static and dynamical consequences of motional 
moduation of the magnetic interaction, we performed a series of quantum 
Monte Carlo (QMC) calculations. By using the motional parameters for 
La$_2$Cu$^{16}$O$_4$ at 15 and 290~K, we generated lattices with a distribution 
of coupling values to represent snapshots of the ionic motion. The staggered 
magnetization in the ground state was then determined by finite-size 
extrapolation of $L \times L$ lattices using 10 realizations of the ionic 
positions for $L$ = 16, 32, and 64 and five realizations for $L = 128$. 
The result for uniform couplings, $m_s = 0.3072 \pm 0.0007$, is in good 
agreement with literature results.\cite{rms} The results for the lattice 
of interaction strengths modulated by the positional distribution at 15 K 
is $m_s = 0.3045 \pm 0.0002$, a relative reduction somewhat smaller than 
that in $\langle J \rangle$. While the extent to which a frozen-distribution 
approach can capture correctly the effects of ionic motion remains an open 
question, we expect that it is appropriate for a static property such as 
$m_s$, which is slow compared to the time scale of the ionic motion (which 
in turn is slow on the scale of the electronic motion). 

As an example of a dynamical quantity, we have investigated the 
consequences of ionic motion for the high-energy spin excitations in 
La$_2$CuO$_4$, which occur around 300~meV. Because this energy lies well 
above all the phonon frequencies, any motional effects on the spin response 
should be independent of specific lattice vibrations. We focus on the 
antiferromagnetic zone-boundary point, ${\bf q} = (\pi/2,\pi/2)$, in 
order to separate motional fluctuation effects from the intrinsic 
quantum effects observed around the $(\pi,0)$ point in 2D Heisenberg 
antiferromagnets.\cite{christensen2004,christensen2007,tsyrulin2010} We 
calculated the transverse dynamical structure factor on a 16$\times$16 lattice 
with periodic boundary conditions, using the stochastic series expansion 
quantum Monte Carlo technique with directed loop updates.\cite{syljuasen2002} 
To determine the transverse component, a small field $h_z = 0.01 J_0$ was 
imposed to break the spin rotational symmetry. The dynamical quantities 
were extracted from the imaginary-time Monte Carlo data using a stochastic 
analytic continuation technique.\cite{sandvik1998,syljuasen2008} 

\begin{figure}[bth]
\noindent\includegraphics[width=0.48\textwidth]{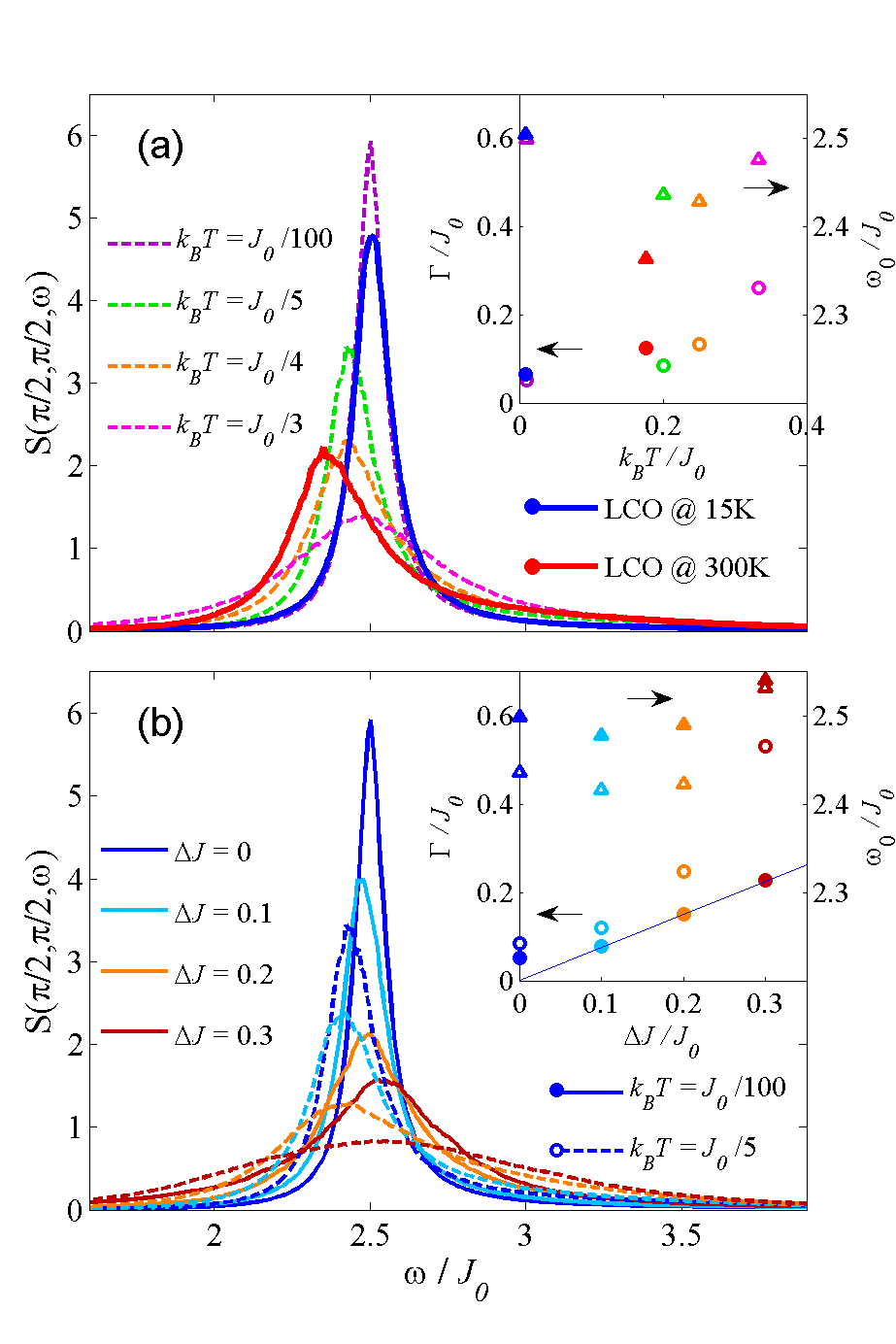}
\caption{(Color online) Energy dependence of the transverse dynamical 
structure factor $S({\bf q},\omega)$ at ${\bf q} = (\pi/2,\pi/2)$, computed 
by quantum Monte Carlo simulations. (a) Calculations for La$_2$CuO$_4$ using 
the $J$ distributions appropriate for the measured ionic motion at 15~K and 
290~K (solid lines). For comparison we show as dashed lines the results 
expected from a purely thermal broadening, i.e.~with constant $J = J_0$, 
for chosen values of $J_0/k_BT$. Inset: temperature dependence of peak 
energy $\omega_0/J_0$ (triangles) and peak width $\Gamma/J_0$ (circles) for 
La$_2$CuO$_4$ at both experimental temperatures (solid symbols) and for 
uniform $J$ with thermal broadening (open symbols). (b) Calculations 
made using Gaussian distributions of $J$ values with increasing standard 
deviations $\Delta J$ for $k_BT = J_0/100$ (solid lines) and for $k_BT = 
J_0/5$ (dashed lines). Inset: dependence of $\Gamma$ (circles) and $\omega_0$ 
(triangles) for $k_BT = J_0/100$ (solid symbols) and for $k_BT = J_0/5$ (open 
symbols). The line indicates the linear relationship $\Gamma = 0.75 
\Delta J$ for $k_BT = J_0/100$.} 
\label{fig:qmc}
\end{figure}

Figure \ref{fig:qmc}(a) compares the temperature dependence of the line 
shape obtained for a homogeneous lattice, meaning with constant values 
$J = J_0$ on every bond, to that obtained using $J$ values distributed 
according to the motional parameters we have deduced for La$_2$Cu$^{16}$O$_4$ 
at 15 K ($k_BT \simeq 0.01 J_0$) and at 290~K ($k_BT \simeq 0.2 J_0$).
We characterize the line shape by the line width $\Gamma$, which we deduce 
from the half-width at half maximum height (HWHM) on the low-energy side of 
the peak in the dynamical structure factor, because the high-energy side of 
the spectrum is extended by continuum states. The finite line width, $\Gamma_i
 = 0.051 J$, for a constant $J$ at low temperatures is a consequence of 
these higher-energy continua combined with the limitations inherent in 
analytic continuation of QMC data.\cite{syljuasen2000} However, we will 
show below that majority of the broadening we calculate for La$_2$CuO$_4$ 
at 15~K, $\Gamma = 0.064 J$, is in fact a consequence of the zero-point 
fluctuations in $J$ and is not due to this intrinsic QMC line width. On
raising the temperature, we find that the combined effects of thermal spin 
fluctuations and ionic motion lead to a very strong broadening $\Gamma = 
0.124J$ at 290~K, which is 50\% wider than the value $\Gamma = 0.084J$ 
obtained for the uniform lattice at $k_BT = 0.2 J_0$ [Figure \ref{fig:qmc}(a)]. 

To further elucidate the effect of a distribution of $J$ values on the 
excitation line width, and of how this effect combines with thermal spin 
fluctuations, we show in Fig.~\ref{fig:qmc}(b) calculations performed at 
$k_BT = J_0/100$ and $k_BT = J_0/5$ using Gaussian distributions of $J$ with 
standard deviations from 0 to $0.3J_0$. At low temperature, this effective 
fluctuation of $J$ causes a linear increase in the excitation line width, 
$\Gamma = 0.75 \Delta J$ [inset, Fig.~\ref{fig:qmc}(b)]. This is the 
observation allowing us to disentangle a broadening of $S({\bf q},\omega)$ 
induced by ionic motion ($J$-fluctuation) from the intrinsic low-temperature 
QMC line width, $\Gamma_i$, computed with uniform $J$ values. Indeed, taking 
$\Delta J = 0.099 J_0^r$ at 15~K from Table IV, a linear interpolation yields 
$\Gamma = 0.074 J$, a value close to that simulated directly for La$_2$CuO$_4$ 
at this temperature. These results demonstrate that this calculated line width 
arises primarily from ionic motion, reflecting little or no influence from 
$\Gamma_i$ when $\Gamma > \Gamma_i$. Adding thermal spin fluctuations 
corresponding to $k_BT = J_0/5$ has only a small effect on the uniform-$J$ 
lattice, but strongly enhances the broadening on the distributed-$J$ lattices. 
By contrast, the effect of thermal fluctuations on the peak position seems 
to be only an additive constant; fluctuations of $J$ lead to rather small 
shifts in peak position, with a minimum around $\Delta J \sim 0.15J$, which 
happens to correspond to La$_2$CuO$_4$ at 290~K, while the softening caused 
by thermal spin fluctuations is almost constant (2.5\%) for $0 \le \Delta J 
\le 0.2J_0$. The width and apparent hardening we find for $\Delta J = 0.3 J_0$ 
at $k_BT = J_0/5$ is probably the result of the broadening magnon peak merging 
with the higher-energy continuum in the analytic continuation.

From these results we may conclude that the distribution of $J$ 
values at 15~K causes a line broadening of order 12\% in the FWHM, which 
is approximately $5\%$ of the spin-wave energy at ${\bf q} = (\pi/2,\pi/2)$. 
At 290~K, this broadening doubles to a FWHM of order $10\%$ of the spin-wave 
energy, accompanied by a sizeable softening (below). These levels of broadening 
should be readily detectable at time-of-flight neutron spectrometers such as 
MAPS at the ISIS spallation source (UK), which can reach 2$-$5\% energy 
resolutions (FWHM) by employing higher-resolution instrument configurations 
than those used in previous experiments.\cite{coldea,headings2010} We also 
draw attention to the fact that ionic motion leads to significantly larger 
softening of the peak position (5.5\% at 290~K) than do thermal spin 
fluctuations alone (2.5\%). Indeed, the values reported\cite{coldea} for 
La$_2$CuO$_4$, $J = 146.3$~meV at 10~K and $J = 138.3$~meV at 300~K, differ 
by exactly 5.5\%. However, a more detailed analysis taking into account the 
further-neighbor interactions\cite{guarise2010,dallapiazza2012} would be 
required for such a comparison to be conclusive.

Finally, we note that studying the distributions of $t$ and $J$ in the form 
of the histograms in Fig.~\ref{histograms}, and characterizing these by their 
mean and width, neglects bond correlations arising from the motion of the Cu 
ions. It is clear that one Cu atom moving towards $+x$ increases the value of 
$J_{\langle i,i+x \rangle}$ while simultaneously reducing $J_{\langle i-x,i \rangle}$, 
and similar correlations can be expected between the $x$- and $y$-directions. 
Because we use a finite lattice of frozen $J$ values, our calculations of the 
dynamical structure factor include these bond correlations. Their effects 
can be gauged by comparing the results in Fig.~\ref{fig:qmc}(a) with those 
obtained from a randomized set of lattice bond strengths. We find that this 
procedure gives barely discernible increases in the line widths (data not 
shown), and hence that comparisons with a purely Gaussian broadening are fully 
justified. We expect that this very small narrowing represents the full extent 
of correlation effects at the temperatures of interest here, where the Born 
approximation remains intact and ionic motion is essentially incoherent. 
However, at higher lattice energies, a deeper analysis of their consequences 
for electronic and magnetic properties would include phonon-mediated hopping 
and the possible enhancement of polaronic physics, topics lying beyond the 
scope of the current manuscript.

\section{Summary}

We have studied the quantum and thermal contributions to ionic motion in 
the antiferromagnetic insulator La$_2$CuO$_{4}$ by means of high-resolution 
neutron diffraction experiments. We found that the anisotropic deformation 
of the ellipsoids of ionic motion is different for in-plane and out-of-plane 
ions. The elongation is most pronounced along the $z$-direction for Cu and O1 
ions, while in-plane O motion reflects its bond orientation; by contrast, for 
the La and O2 ions one sees at 290 K that in-plane motion is stronger. 
Zero-point, or quantum, fluctuations account for more than half of the 
ionic motion observed at 290 K. 

By using samples of 100\% $^{16}$O and 78\% $^{18}$O substitition, we 
investigated the influence of the oxygen isotope on the structural 
and thermal parameters of La$_2$CuO$_4$. We found a nonvanishing negative 
OIE on the lattice parameters of approximately 0.01\%. The negative OIE on 
$c$ is in qualitative agreement with the positive OIE we measured in the 
N\'eel temperature $T_N$. The OIE on the Cu--O1 bond length is also negative. 
Further, our results demonstrate a considerable OIE on the zero-point motion, 
which is positive for the La and Cu atoms but negative for the O1 and O2 atoms. 
The preferred directions of ionic motion are identical for the $^{16}$O and 
$^{18}$O samples.

Working within the Born approximation, we use our detailed structural data 
to perform a theoretical study of how the zero-point and thermal motion of 
Cu and O ions will affect the electronic and magnetic properties of cuprates 
through the effective hopping integral $t$ and the antiferromagnetic exchange 
parameter $J$. By modelling the spatial distribution of ionic positions, we 
demonstrate that $J$ undergoes a Gaussian broadening, which is significant 
(exceeding 10\%) even at low temperatures and can be of order 20\% at room 
temperature. This broadening shows a measurable positive OIE as a consequence 
of heavier O ions allowing enhanced Cu ion motion. 

Our results suggest that this broadening cannot be neglected in the detailed 
theoretical modelling of cuprate systems. To illustrate this we compute one 
static quantity and one dynamical one. The staggered magnetization is reduced 
only weakly by motional renormalization, changing less than the shift in the 
average of the $J$ distribution and thereby showing only percent-level effects. 
However, the transverse dynamical structure factor shows line broadening and 
peak intensity loss at the 10\% level due to zero-point ionic motion, and very 
dramatic suppression and broadening due to thermal effects; at 290~K, the 
combination of thermal motional and spin fluctations can lead to changes 
by a factor of two. These important renormalization effects, obtained 
within the approximation of incoherent ionic motion, indicate that models 
based on a static lattice may be insufficient for a full description of the 
electronic and magnetic properties of cuprates, and by extension of other 
similar transition-metal compounds.

\section{Acknowledgments}

Neutron powder diffraction results are based on experiments performed at 
the Swiss spallation neutron source SINQ, at the Paul Scherrer Institute, 
Villigen, Switzerland. We thank H.~Keller and B.~Batlogg for fruitful 
discussions, B.~R\"ossner for helpful contributions, and S. M. Hayden for 
providing information on the resolution of previous neutron scattering 
experiments. This work was supported by the NCCR MaNEP and the Synergia 
network on Mott Physics Beyond the Heisenberg Model of the Swiss NSF; by 
the NSF of China under Grant 11174365 and the National Basic Research 
Program of the Chinese MoST under Grant 2012CB921704; and by the Norwegian 
Research Council under NOTUR Grant nn4563k for using the Abel computer cluster.


\begin{thebibliography}{21}
\expandafter\ifx\csname natexlab\endcsname\relax\def\natexlab#1{#1}\fi
\expandafter\ifx\csname bibnamefont\endcsname\relax
  \def\bibnamefont#1{#1}\fi
\expandafter\ifx\csname bibfnamefont\endcsname\relax
  \def\bibfnamefont#1{#1}\fi
\expandafter\ifx\csname citenamefont\endcsname\relax
  \def\citenamefont#1{#1}\fi
\expandafter\ifx\csname url\endcsname\relax
  \def\url#1{\texttt{#1}}\fi
\expandafter\ifx\csname urlprefix\endcsname\relax\def\urlprefix{URL }\fi
\providecommand{\bibinfo}[2]{#2}
\providecommand{\eprint}[2][]{\url{#2}}

\bibitem{plakida} N. M. Plakida, {\sl High-Temperature Cuprate Superconductors: 
Experiment, Theory, and Applications} (Springer, Heidelberg, 2010).

\bibitem[{\citenamefont{Kastner et~al.}(1998)\citenamefont{Kastner, Birgeneau,
  Shirane, and Endoh}}]{kastner}
\bibinfo{author}{\bibfnamefont{M.~A.} \bibnamefont{Kastner}},
  \bibinfo{author}{\bibfnamefont{R.~J.} \bibnamefont{Birgeneau}},
  \bibinfo{author}{\bibfnamefont{G.}~\bibnamefont{Shirane}}, \bibnamefont{and}
  \bibinfo{author}{\bibfnamefont{Y.}~\bibnamefont{Endoh}},
  \bibinfo{journal}{Rev. Mod. Phys.} \textbf{\bibinfo{volume}{70}},
  \bibinfo{pages}{897} (\bibinfo{year}{1998}).

\bibitem{huse} D. A. Huse, Phys. Rev. B {\bf 37}, 2380 (1988).

\bibitem[{\citenamefont{Coldea et~al.}(2001)\citenamefont{Coldea, Hayden,
  Aeppli, Perring, Frost, Mason, Cheong, and Fisk}}]{coldea}
\bibinfo{author}{\bibfnamefont{R.}~\bibnamefont{Coldea}},
  \bibinfo{author}{\bibfnamefont{S.~M.} \bibnamefont{Hayden}},
  \bibinfo{author}{\bibfnamefont{G.}~\bibnamefont{Aeppli}},
  \bibinfo{author}{\bibfnamefont{T.~G.} \bibnamefont{Perring}},
  \bibinfo{author}{\bibfnamefont{C.~D.} \bibnamefont{Frost}},
  \bibinfo{author}{\bibfnamefont{T.~E.} \bibnamefont{Mason}},
  \bibinfo{author}{\bibfnamefont{S.-W.} \bibnamefont{Cheong}},
  \bibnamefont{and} \bibinfo{author}{\bibfnamefont{Z.}~\bibnamefont{Fisk}},
  \bibinfo{journal}{Phys. Rev. Lett.} \textbf{\bibinfo{volume}{86}},
  \bibinfo{pages}{5377} (\bibinfo{year}{2001}).

\bibitem[{\citenamefont{R\o{}nnow et~al.}(2001)\citenamefont{R\o{}nnow,
  McMorrow, Coldea, Harrison, Youngson, Perring, Aeppli, Sylju\aa{}sen,
  Lefmann, and Rischel}}]{ronnow2001}
\bibinfo{author}{\bibfnamefont{H.~M.} \bibnamefont{R\o{}nnow}},
  \bibinfo{author}{\bibfnamefont{D.~F.} \bibnamefont{McMorrow}},
  \bibinfo{author}{\bibfnamefont{R.}~\bibnamefont{Coldea}},
  \bibinfo{author}{\bibfnamefont{A.}~\bibnamefont{Harrison}},
  \bibinfo{author}{\bibfnamefont{I.~D.} \bibnamefont{Youngson}},
  \bibinfo{author}{\bibfnamefont{T.~G.} \bibnamefont{Perring}},
  \bibinfo{author}{\bibfnamefont{G.}~\bibnamefont{Aeppli}},
  \bibinfo{author}{\bibfnamefont{O.}~\bibnamefont{Sylju\aa{}sen}},
  \bibinfo{author}{\bibfnamefont{K.}~\bibnamefont{Lefmann}}, \bibnamefont{and}
  \bibinfo{author}{\bibfnamefont{C.}~\bibnamefont{Rischel}},
  \bibinfo{journal}{Phys. Rev. Lett.} \textbf{\bibinfo{volume}{87}},
  \bibinfo{pages}{037202} (\bibinfo{year}{2001}).

\bibitem[{\citenamefont{Anderson}(1987)}]{Anderson3}
\bibinfo{author}{\bibfnamefont{P.~W.} \bibnamefont{Anderson}},
  \bibinfo{journal}{Science} \textbf{\bibinfo{volume}{235}},
  \bibinfo{pages}{1196} (\bibinfo{year}{1987}).

\bibitem[{\citenamefont{Kohanoff et~al.}(1992)\citenamefont{Kohanoff, Andreoni,
  and Parrinello}}]{kohanoff}
\bibinfo{author}{\bibfnamefont{J.}~\bibnamefont{Kohanoff}},
  \bibinfo{author}{\bibfnamefont{W.}~\bibnamefont{Andreoni}}, \bibnamefont{and}
  \bibinfo{author}{\bibfnamefont{M.}~\bibnamefont{Parrinello}},
  \bibinfo{journal}{Phys. Rev. B} \textbf{\bibinfo{volume}{46}},
  \bibinfo{pages}{4371} (\bibinfo{year}{1992}).

\bibitem{maxwell} E. Maxwell, Phys. Rev. {\bf 78}, 477 (1950). 

\bibitem{reynolds} C. A. Reynolds, B. Serin, W. H. Wright, and L. B. Nesbitt, 
{\bf 78}, 487 (1950). 

\bibitem{froehlich} H. Fr\"ohlich, Proc. Phys. Soc. {\bf A63}, 778 (1950). 

\bibitem[{\citenamefont{Hafliger et~al.}(2006)\citenamefont{Hafliger,
  Podlesnyak, Conder, Pomjakushina, and Furrer}}]{petra}
\bibinfo{author}{\bibfnamefont{P.~S.} \bibnamefont{H\"afliger}},
  \bibinfo{author}{\bibfnamefont{A.}~\bibnamefont{Podlesnyak}},
  \bibinfo{author}{\bibfnamefont{K.}~\bibnamefont{Conder}},
  \bibinfo{author}{\bibfnamefont{E.}~\bibnamefont{Pomjakushina}},
  \bibnamefont{and} \bibinfo{author}{\bibfnamefont{A.}~\bibnamefont{Furrer}},
  \bibinfo{journal}{Phys. Rev. B}
  \textbf{\bibinfo{volume}{74}}, \bibinfo{eid}{184520}
   (\bibinfo{year}{2006}),

\bibitem{keller} H. Keller in \emph{Superconductivity in Complex Systems: 
Structure and Bonding}, eds. K. A. M\"{u}ller and A. Bussmann-Holder, 
vol. {\bf 114}, p.~143 (Springer, Berlin, 2005). 

\bibitem[{\citenamefont{Khasanov et~al.}(2008)\citenamefont{Khasanov,
  Shengelaya, Castro, Morenzoni, Maisuradze, Savic, Conder, Pomjakushina,
  Bussmann-Holder, and Keller}}]{rustem}
\bibinfo{author}{\bibfnamefont{R.}~\bibnamefont{Khasanov}},
  \bibinfo{author}{\bibfnamefont{A.}~\bibnamefont{Shengelaya}},
  \bibinfo{author}{\bibfnamefont{D.} \bibnamefont{Di Castro}},
  \bibinfo{author}{\bibfnamefont{E.}~\bibnamefont{Morenzoni}},
  \bibinfo{author}{\bibfnamefont{A.}~\bibnamefont{Maisuradze}},
  \bibinfo{author}{\bibfnamefont{I.~M.} \bibnamefont{Savic}},
  \bibinfo{author}{\bibfnamefont{K.}~\bibnamefont{Conder}},
  \bibinfo{author}{\bibfnamefont{E.}~\bibnamefont{Pomjakushina}},
  \bibinfo{author}{\bibfnamefont{A.}~\bibnamefont{Bussmann-Holder}},
  \bibnamefont{and} \bibinfo{author}{\bibfnamefont{H.}~\bibnamefont{Keller}},
  \bibinfo{journal}{Phys. Rev. Lett.} \textbf{\bibinfo{volume}{101}},
  \bibinfo{pages}{077001} (\bibinfo{year}{2008}).

\bibitem[{\citenamefont{A.~Shengelaya and Keller}(1999)}]{shengelaya}
\bibinfo{author}{ \bibnamefont{A.~Shengelaya},
  \bibfnamefont{G. M.~Zhao}}, C. M. Aegerter, K. Conder, I. M. Savic, and
  \bibinfo{author}{\bibfnamefont{H.}~\bibnamefont{Keller}},
  \bibinfo{journal}{Phys. Rev. Lett.} \textbf{\bibinfo{volume}{83}},
  \bibinfo{pages}{5142} (\bibinfo{year}{1999}).

\bibitem[{\citenamefont{Bussmann-Holder and Keller}(2007)}]{bussmann}
\bibinfo{author}{\bibfnamefont{A.}~\bibnamefont{Bussmann-Holder}}
  \bibnamefont{and} \bibinfo{author}{\bibfnamefont{H.}~\bibnamefont{Keller}},
  \bibinfo{journal}{in {\emph Polarons in Advanced Materials}, ed. A. S. 
   Alexandrov, p.~599 (Springer, Dordrecht \& Canopus Publishing, Bristol, 
   2007)} 

\bibitem[{\citenamefont{Zhao et~al.}(1994)\citenamefont{Zhao, Singh, and
  Morris}}]{zhao}
\bibinfo{author}{\bibfnamefont{G. M.} \bibnamefont{Zhao}},
  \bibinfo{author}{\bibfnamefont{K.~K.} \bibnamefont{Singh}}, \bibnamefont{and}
  \bibinfo{author}{\bibfnamefont{D.~E.} \bibnamefont{Morris}},
  \bibinfo{journal}{Phys. Rev. B} \textbf{\bibinfo{volume}{50}},
  \bibinfo{pages}{4112} (\bibinfo{year}{1994}).

\bibitem[{\citenamefont{Hanzawa}(1995)}]{Hanzawa}
\bibinfo{author}{\bibfnamefont{K.}~\bibnamefont{Hanzawa}}, \bibinfo{journal}{J.
  Soc. Phys. Jpn} \textbf{\bibinfo{volume}{64}}, \bibinfo{pages}{4856}
  (\bibinfo{year}{1995}).

\bibitem[{\citenamefont{Conder}(2001)}]{conder}
\bibinfo{author}{\bibfnamefont{K.}~\bibnamefont{Conder}},
  \bibinfo{journal}{Mater. Sci. Eng.} \textbf{\bibinfo{volume}{R 32}},
  \bibinfo{pages}{41} (\bibinfo{year}{2001}).

\bibitem[{\citenamefont{K.~Conder}(2002)}]{conder2}
\bibinfo{author}{\bibfnamefont{K.~Conder}}, 
\bibinfo{author}{\bibnamefont{G. M. Zhao}}, \bibnamefont{and} 
\bibinfo{author}{\bibnamefont{R. Khasanov}}, \bibinfo{journal}{Phys. Rev. B}
  \textbf{\bibinfo{volume}{66}}, \bibinfo{pages}{212409}
  (\bibinfo{year}{2002}).

\bibitem[{\citenamefont{Fischer}(2000)}]{hrpt}
\bibinfo{author}{\bibfnamefont{P.}~\bibnamefont{Fischer}},
  \bibinfo{journal}{Physica B} \textbf{\bibinfo{volume}{276-278}},
  \bibinfo{pages}{146} (\bibinfo{year}{2000}).

\bibitem[{\citenamefont{Fischer}(1997)}]{sinq}
\bibinfo{author}{\bibfnamefont{W.~E.} \bibnamefont{Fischer}},
  \bibinfo{journal}{Physica B} \textbf{\bibinfo{volume}{234-236}},
  \bibinfo{pages}{1202} (\bibinfo{year}{1997}).

\bibitem[{\citenamefont{Radaelli et~al.}(1994)\citenamefont{Radaelli, Hinks,
  Mitchell, Hunter, Wagner, Dabrowski, Vandervoort, Viswanathan, and
  Jorgensen}}]{radaelli}
\bibinfo{author}{\bibfnamefont{P.~G.} \bibnamefont{Radaelli}},
  \bibinfo{author}{\bibfnamefont{D.~G.} \bibnamefont{Hinks}},
  \bibinfo{author}{\bibfnamefont{A.~W.} \bibnamefont{Mitchell}},
  \bibinfo{author}{\bibfnamefont{B.~A.} \bibnamefont{Hunter}},
  \bibinfo{author}{\bibfnamefont{J.~L.} \bibnamefont{Wagner}},
  \bibinfo{author}{\bibfnamefont{B.}~\bibnamefont{Dabrowski}},
  \bibinfo{author}{\bibfnamefont{K.~G.} \bibnamefont{Vandervoort}},
  \bibinfo{author}{\bibfnamefont{H.~K.} \bibnamefont{Viswanathan}},
  \bibnamefont{and} \bibinfo{author}{\bibfnamefont{J.~D.}
  \bibnamefont{Jorgensen}}, \bibinfo{journal}{Phys. Rev. B}
  \textbf{\bibinfo{volume}{49}}, \bibinfo{pages}{4163} (\bibinfo{year}{1994}).

\bibitem{rrc}
J. Rodr\'iguez-Carvajal, Physica B {\bf 192}, 55 (1993).

\bibitem[{\citenamefont{Jorgensen et~al.}(1988)\citenamefont{Jorgensen,
  Dabrowski, Pei, Hinks, Soderholm, Morosin, Schirber, Venturini, and
  Ginley}}]{jorgensen}
\bibinfo{author}{\bibfnamefont{J.~D.} \bibnamefont{Jorgensen}},
  \bibinfo{author}{\bibfnamefont{B.}~\bibnamefont{Dabrowski}},
  \bibinfo{author}{\bibfnamefont{S.}~\bibnamefont{Pei}},
  \bibinfo{author}{\bibfnamefont{D.~G.} \bibnamefont{Hinks}},
  \bibinfo{author}{\bibfnamefont{L.}~\bibnamefont{Soderholm}},
  \bibinfo{author}{\bibfnamefont{B.}~\bibnamefont{Morosin}},
  \bibinfo{author}{\bibfnamefont{J.~E.} \bibnamefont{Schirber}},
  \bibinfo{author}{\bibfnamefont{E.~L.} \bibnamefont{Venturini}},
  \bibnamefont{and} \bibinfo{author}{\bibfnamefont{D.~S.}
  \bibnamefont{Ginley}}, \bibinfo{journal}{Phys. Rev. B}
  \textbf{\bibinfo{volume}{38}}, \bibinfo{pages}{11337} (\bibinfo{year}{1988}).

\bibitem[{\citenamefont{et~al.}(1990)}]{chaillout}
\bibinfo{author}{\bibfnamefont{C.} \bibnamefont{Chaillout}},
\bibinfo{author}{\bibfnamefont{J.} \bibnamefont{Chenavas}},
\bibinfo{author}{\bibfnamefont{S.~W.} \bibnamefont{Cheong}},
\bibinfo{author}{\bibfnamefont{Z.} \bibnamefont{Fisk}},
\bibinfo{author}{\bibfnamefont{M.} \bibnamefont{Marezio}},
\bibinfo{author}{\bibfnamefont{B.} \bibnamefont{Morosin}}, \bibnamefont{and}
\bibinfo{author}{\bibfnamefont{J.~E.} \bibnamefont{Schirber}},
  \bibinfo{journal}{Physica C} \textbf{\bibinfo{volume}{170}},
  \bibinfo{pages}{87} (\bibinfo{year}{1990}).

\bibitem[{\citenamefont{Rosenstock}(1963)}]{rosenstock}
\bibinfo{author}{\bibfnamefont{H.~B.} \bibnamefont{Rosenstock}},
  \bibinfo{journal}{Phys. Rev.} \textbf{\bibinfo{volume}{129}},
  \bibinfo{pages}{1959} (\bibinfo{year}{1963}).

\bibitem[{\citenamefont{B\"oni et~al.}(1988)\citenamefont{B\"oni, Axe, Shirane,
  Birgeneau, Gabbe, Jenssen, Kastner, Peters, Picone, and Thurston}}]{boeni}
\bibinfo{author}{\bibfnamefont{P.}~\bibnamefont{B\"oni}},
  \bibinfo{author}{\bibfnamefont{J.~D.} \bibnamefont{Axe}},
  \bibinfo{author}{\bibfnamefont{G.}~\bibnamefont{Shirane}},
  \bibinfo{author}{\bibfnamefont{R.~J.} \bibnamefont{Birgeneau}},
  \bibinfo{author}{\bibfnamefont{D.~R.} \bibnamefont{Gabbe}},
  \bibinfo{author}{\bibfnamefont{H.~P.} \bibnamefont{Jenssen}},
  \bibinfo{author}{\bibfnamefont{M.~A.} \bibnamefont{Kastner}},
  \bibinfo{author}{\bibfnamefont{C.~J.} \bibnamefont{Peters}},
  \bibinfo{author}{\bibfnamefont{P.~J.} \bibnamefont{Picone}},
  \bibnamefont{and} \bibinfo{author}{\bibfnamefont{T.~R.}
  \bibnamefont{Thurston}}, \bibinfo{journal}{Phys. Rev. B}
  \textbf{\bibinfo{volume}{38}}, \bibinfo{pages}{185} (\bibinfo{year}{1988}).

\bibitem{rzr}
F.-C. Zhang and T. M. Rice, Phys. Rev. B {\bf 37}, 3759 (1988). 

\bibitem{rej}
H. Eskes and J. H. Jefferson, 
Phys. Rev. B {\bf 48}, 9788 (1993).

\bibitem{raea}
M. C. Aronson, S. B. Dierker, B. S. Dennis, S.-W. Cheong, and Z. Fisk,
Phys. Rev. B {\bf 44}, 4657 (1991).

\bibitem{rld}
K.-P. Bohnen, R. Heid, and M. Krauss, Europhys. Lett. {\bf 64}, 104 (2003).

\bibitem{phononenergy}
A. Lanzara, P. V. Bogdanov, X. J. Zhou, S. A. Kellar, D. L. Feng, E. D. Lu, 
T. Yoshida, H. Eisaki, A. Fujimori, K. Kishio, J. I. Shimoyama, T. Noda, 
S. Uchida, Z. Hussain, and Z. X. Shen, Nature {\bf 412}, 510 (2001).

\bibitem{rb}
S. Bramwell, J. Phys. Condens. Matter {\bf 2}, 7257 (1990).

\bibitem{rhb}
W. A. Harrison, {\sl Orbital Overlap and the Chemical Bond} (Wiley, 
New York, 1980).

\bibitem{ronnow1999}
H. M. R\o{}nnow, D. F. McMorrow, and A. Harrison, Phys. Rev. Lett. {\bf 82}, 
3152 (1999). 

\bibitem{rie1} 
G. M. Zhao, H. Keller, and K. Conder, J. Phys. Condens. Matt. {\bf 13}, 
R569 (2001), and references therein.

\bibitem{rie2} 
C. Bernhard, T. Holden, A. V. Boris, N. N. Kovaleva, A. V. Pimenov, J. 
Humlicek, C. Ulrich, C. T. Lin, and J. L. Tallon, Phys. Rev. B {\bf 69}, 
052502 (2004).

\bibitem{rie3} 
G.-H. Gweon, T. Sasagawa, S. Y. Zhou, J. Graf, H. Takagi, D.-H. Lee, and 
A. Lanzara, Nature {\bf 430}, 187 (2004).

\bibitem{rms}
D. A. Huse, Phys. Rev. B {\bf 37}, 2380 (1988). 

\bibitem{christensen2004}
N. B. Christensen, D. F. McMorrow, H. M. R\o{}nnow, A. Harrison, 
T. G. Perring, and R. Coldea, J. Mag. Mag. Mat. {\bf 272-274}, 896 (2004).

\bibitem{christensen2007}
N. B. Christensen, H. M. R\o{}nnow, D. F. McMorrow, A. Harrison, T. G. Perring,
M. Enderle, R. Coldea, L. P. Regnault, and G. Aeppli, Proc. Natl. Acad. Sci. 
{\bf 104}, 15264 (2007).

\bibitem{tsyrulin2010}
N. Tsyrulin, F. Xiao, A. Schneidewind, P. Link, H. M. R\o{}nnow, J. Gavilano, 
C. P. Landee, M. M. Turnbull, and M. Kenzelmann, Phys. Rev. B {\bf 81}, 134409 
(2010).

\bibitem{syljuasen2002}
O.~F. Sylju{\aa}sen and A.~W. Sandvik, Phys. Rev. E. {\bf 66}, 046701 (2002).

\bibitem{sandvik1998}
A.~W. Sandvik, 
Phys. Rev. B {\bf 57},  10287 (1998).

\bibitem{syljuasen2008}
O.~F. Sylju{\aa}sen, Phys. Rev. B {\bf 78}, 174429 (2008).

\bibitem{syljuasen2000}
O. F. Sylju{\aa}sen and H. M. R\o{}nnow, J. Phys. C {\bf 12}, L405 (2000).

\bibitem{ronnow2001b}
H. M. R\o{}nnow, D. F. McMorrow, A. Harrison, I. D. Youngson, R. Coldea, 
T. G. Perring, G. Aeppli, and O. Sylju{\aa}sen, J. Mag. Mag. Mater. {\bf 236}, 
4 (2001).

\bibitem{headings2010}
N. S. Headings, S. M. Hayden, R. Coldea, and T. G. Perring, Phys. Rev. Lett. 
{\bf 105}, 247001 (2010).

\bibitem{guarise2010}
M. Guarise, B. Dalla Piazza, M. Moretti Sala, G. Ghiringhelli, L. Braicovich, 
H. Berger, J. N. Hancock, D. van der Marel, T. Schmitt, V. N. Strocov, L. J. 
P. Ament, J. van den Brink, P.-H. Lin, P. Xu, H. M. R\o{}nnow, and M. Grioni, 
Phys. Rev. Lett. {\bf 105}, 157006 (2010).

\bibitem{dallapiazza2012}
B. Dalla Piazza, M. Mourigal, M. Guarise, H. Berger, T. Schmitt, K. J. Zhou, 
M. Grioni, and H. M. R\o{}nnow, Phys. Rev. B {\bf 85}, 100508 (2012).

\end{thebibliography}
\end{document}